\documentclass{aa}

\usepackage{txfonts,graphicx,rotating,multirow}

\begin{document}


\title{Evolution of interstellar dust and stardust in the solar neighbourhood}


\author{Svitlana Zhukovska\inst{1} \and Hans-Peter Gail\inst{1} \and Mario
Trieloff\inst{2}
}

\institute{ZAH, Institut f\"ur Theoretische Astrophysik, Albert-\"Uberle-Str. 2,
           D-69120 Heidelberg, Germany
      \and
           Mineralogical Institute, University of Heidelberg, Im Neuenheimer
           Feld 236, D-69121 Heidelberg, Germany 
  }

\offprints{H.-P. Gail}

\date{Received date ; accepted date}

\abstract{}
{The abundance evolution of interstellar dust species originating from stellar
sources and from condensation in molecular clouds in the local interstellar
medium of the Milky Way is studied and the input of dust material to the
Solar System is determined.}
{A one-zone chemical evolution model of the Milky Way for the elemental
composition of the disk combined with an evolution model for its interstellar
dust component similar to that of Dwek (1998) is developed. The dust model considers dust-mass return from AGB stars as calculated from synthetic AGB models combined with models for dust condensation in stellar outflows. Supernova dust formation is included in a simple parameterized form which is gauged by observed abundances of presolar dust grains with supernova origin. For dust growth in the ISM a simple method is developed for coupling this with disk and dust evolution models.}
{The time evolution of the abundance of the following dust species is followed in the model: silicate, carbon, silicon carbide, and iron dust from AGB stars and from SNe as well as silicate, carbon, and iron dust grown in molecular clouds. It is shown that the interstellar dust population is dominated by dust accreted in molecular clouds; most of the dust material entering the Solar System at its formation does not show isotopic abundance anomalies of the refractory elements, i.e., inconspicuous isotopic abundances do not point to a Solar System origin of dust grains. The observed abundance ratios of presolar dust grains formed in SN ejecta and in AGB star outflows requires that for the ejecta from SNe the fraction of refractory elements condensed into dust is 0.15 for carbon dust and is  quite small ($\sim10^{-4}$) for other dust species.}
{}


\maketitle

\titlerunning{Presolar grains from AGB stars}

\section{Introduction}

In this paper we intend to study the population of dust grains in the
interstellar medium of the Milky Way, its composition and evolution, and the
input of such grains into the Solar System.

Part of these grains are formed in stellar ejecta or stellar winds of highly
evolved stars from the refractory elements therein and the resulting
gas-dust-mixture is ultimately mixed with the general interstellar matter. But
generally only some fraction of the refractory elements in the ejecta is really
condensed into solid phases; a big mass fraction of the refractory elements
returned to the ISM -- or sometimes even most of it -- stays in the gas phase.
Turbulent mixing in the ISM rapidly intermingles the material that is ejected
by the many different stellar sources. While all the dust particles from stellar
sources, called \emph{stardust} particles, retain their peculiar isotopic
compositions of a number of elements indicative for their formation sites, in
the gas phase the material from all sources is mixed together and the resulting
isotopic composition of the mix is different from that of the stardust particles.
The refractory elements in the ISM gas therefore have different isotopic
compositions than the same elements found in stardust grains.

\begin{table*}
\caption{Comparison of some observed properties of the galactic disk in the
solar neighbourhood with model results}
\label{TabResults}
\centering
\begin{tabular}{l@{\hspace{1.0cm}}l@{\hspace{1.0cm}}l@{\hspace{1.0cm}}l}
\hline\hline
\noalign{\smallskip}
 Observable & Model & Observed & Reference \\
\noalign{\smallskip}
\hline
\noalign{\smallskip}
Total surface density $\Sigma_{\rm tot}$ [${\rm M_{\sun} pc^{-2}}$]    &
  56  &  $50 - 62$             & Holmberg \& Flynn \cite{Hol04}  \\
ISM surface density $\Sigma_{\rm ISM}$ [${\rm M_{\sun} pc^{-2}}$]       &
  9.7 &  $7 - 13$              & Dickey \cite{Dic93}\\
	&      & $13 - 14$       & Olling \& Merrifield \cite{Oll01} \\
Gas fraction $\Sigma_{\rm ISM}/\Sigma_{\rm tot}$                        &
  0.17 & $0.05 - 0.2$          &\\
Surface density of visible stars $\Sigma_*$ [${\rm M_{\sun} pc^{-2}}$]  & 
  38.6 & $30 - 40$             & Gilmore et al. \cite{Gil89}\\
Surface density of stellar remnants [${\rm M_{\sun} pc^{-2}}$]          & 
  7.7 &  $2 - 4$               & Mera et al. \cite{Mer98}\\
Star formation rate {$B$} [${\rm M_{\sun} pc^{-2} Gyr^{-1}}$]           & 
  3.1 & $3.5 - 5$              & Rana \cite{Ran91}\\
SN II rate $R_{\rm SNII}$ [${\rm pc^{-2} Gyr^{-1}}$]                    & 
  0.016 & $0.009 - 0.0326$     & Tammann et al. \cite{Tam94}\\
SN Ia rate $R_{\rm SNIa}$ [${\rm pc^{-2} Gyr^{-1}}$]                    & 
  $0.0024$ & $0.0015 - 0.0109$ &  Tammann et al. \cite{Tam94}\\
Infall rate [${\rm M_{\sun} pc^{-2} Gyr^{-1}}]$                         & 
  1.45     &  $0.5 - 5$        & Braun \& Thilker \cite{Bra04} \\
\noalign{\smallskip}
\hline
\end{tabular}
\end{table*}


Stardust grains are found in the Solar System as a rare fraction of the fine
grained matrix material of meteorites (e.g. Bernatowicz \& Zinner  \cite{Ber97};
Hoppe \cite{Hop04}; Nguyen et al. \cite{Ngu07}). They are identified as such
by the unusual isotopic composition of at least one element which shows that the
grains have condensed from material that contains freshly synthesized nuclei
from stellar burning zones. Laboratory studies have found a big assortment of
such stardust grains, also called presolar dust particles, with a variety of
chemical compositions that can be associated with a variety of stellar sources.
The composition of the grain material indicates two basically different chemical
environments of formation, (1) a carbon rich environment, and (2) an oxygen rich
environment, which yield two completely different groups of mineral compounds:
\begin{enumerate}
\item solid graphitic carbon, diamond, silicon carbide, silicon
nitride, and
\item corundum, hibonite, spinel, magnesium-iron-silicates.

\end{enumerate}
Besides of these main components a number of minor components (e.g. titanium
oxide, solid solutions of titanium carbide with zirconium and molybdenum
carbide, kamacite, and cohenite) have been identified which are so far only
known to exist as inclusions in grains of the major dust components. Additional
components may exist and await identification, in particular since some kinds
of dust may not survive all the stages between the stellar source and the 
final laboratory investigation. The observed isotopic anomalies indicate
essentially two different kinds of stellar sources of the presolar dust grains,
(1) AGB stars and (2) core collapse supernovae.

From the instant of the formation in stellar ejecta on, stardust grains are
subject to destructive processes in the ISM by sputtering and shattering
processes induced by supernova (SN) shocks (cf. Jones et al. \cite{Jon96}, and
references therein). They are finally incorporated into newly formed stars and
their planetary systems after about 2.5 Gyrs residence time in the ISM, which is
also the typical timescale for replenishment of the ISM with new stardust.
Theoretical studies have shown that typical lifetimes against destruction by
SN shocks are of the order of only about 0.5 Gyrs (Jones et al. \cite{Jon96}).
This rather short timescale compared to the timescale for replenishment would
result in a very low dust abundance in the ISM. This, however, is not observed.
Instead one observes a high degree of depletion of the refractory elements in
the gas phase of the ISM (e.g. Savage \& Sembach \cite{Sav96}; Jenkins
\cite{Jen05}), and this clearly requires additional growth processes in the
interstellar medium that tie up the atoms of the refractory elements in dust.
The only possible sites where accretion of gas phase material onto grains may
proceed with a reasonable short timescale are the dense molecular clouds of the
ISM (Draine \cite{Dra90}). Any solid phase material grown in molecular clouds,
the \emph{MC-grown dust}, has isotopic compositions of the refractory elements
different from that found in the stardust particles and may be identified by
this property. Such dust material may be found both as coating of stardust
grains and as separate grains.

Unfortunately, the isotopic composition of the refractory elements in MC-grown 
dust grains incorporated into the Solar System equals the isotopic
composition of the elements in the Solar System, which makes it impossible to
discriminate by laboratory investigations of isotopic abundances of refractory
elements alone between dust formed in the Solar System and MC-grown dust. There
are other indications, however, which point to a presolar origin of some fraction
of the interplanetary dust particles, the GEMS (Bradley \cite{Bra03}; Messenger
et al. \cite{Mes03}), that shows isotopic abundances of refractory elements
corresponding to normal Solar System isotopic abundances, and which therefore are
likely to be MC-grown dust grains. The category of presolar dust grains therefore
includes also the MC-grown dust species which are isotopically inconspicuous, a
property which makes them presently difficult to be identified as of extrasolar
origin.

If we intend to calculate theoretically the abundances of the different
components of the interstellar dust mixture in the Milky Way at the solar cycle
and in particular the composition of the dust mixture from which the Solar
System formed, we have to construct a model of the Milky Way's chemical evolution
that is coupled with a model for the evolution of the dust component of the
interstellar matter. The evolution of the dust component is not independent of
the evolution of the element abundances since the refractory elements forming the
dust are only gradually formed during the course of the chemical evolution of the
galaxy. The model for the dust evolution needs to consider the injection of
stardust into the ISM, the destruction processes of dust in the ISM, and the
growth processes in molecular clouds. Very simple models for the evolution of
the dust content of galaxies have already been constructed (e.g. Lisenfeld \&
Ferrara \cite{Lis98}; Hirashita \cite{Hir00}; Edmunds \cite{Edm01}; Morgan \& 
Edmunds \cite{Mor03}; Inoue \cite{Ino03}), but these are too simplistic to 
allow for a detailed calculation of the composition of the interstellar dust
mixture. Only the method developed by Dwek (\cite{Dwe98}) to integrate the
chemical evolution of the galactic disk and the dust evolution into a common
model is sufficiently detailed to allow a modeling of the complex interplay
between the processes determining the dust evolution and has the potential of
being extensible to even more complex systems. This model is a one-zone model,
i.e., the galactic disk is approximated by a set of independent cylinders with
all physical variables within a cylinder averaged over the vertical direction 
with respect to the disks midplane, and a one-phase model, i.e., one averages the
properties of the ISM over its different phases (cold, warm, and hot; cf.
Tielens \cite{Tie05}). This type of model allows a successful and at the same
time rather easy calculation of some important properties of the Milky Way disk,
in particular of its chemical evolution (cf. Matteuchi \cite{Mat03}). The price
one has to pay for the simplifications is that some processes, in particular
those depending critically on the phase structure of the ISM, cannot be treated
with sufficient accuracy. Nevertheless, the results obtained by Dwek 
(\cite{Dwe98}) show that such a simple model can successfully be used to
calculate the evolution of the interstellar dust. We take the model of Dwek as a
basis for constructing a model which allows to treat a more complex mixture of
stardust and MC-grown dust. 

For the input of stardust from AGB-stars we use our recent results for the dust
production by AGB-stars (Ferrarotti \& Gail \cite{Fer06}), which are somewhat
extended. These tables present rather detailed information on the amount and
composition of stardust formed by AGB-stars. For stardust from SN we follow the
procedure of Dwek (\cite{Dwe98}) and use a simple parameterization for the dust
production rate since no suited other information on dust production is
available. Observations are inconclusive and theory is only in its infancy
(Schneider et al. \cite{Schn04}; Nozawa et al. \cite{Noz03}, and references
therein). We, in turn, try to gain some insight into the dust production 
efficiency of supernovae by comparing our model results with meteoritic
abundances of stardust from SNe. The dust growth in molecular clouds is
treated in more detail as in the model of Dwek (\cite{Dwe98}) since we intend
to discriminate in the model between stardust and MC-grown dust. In principle
it would also be necessary to consider that according to observations of element
depletion in the ISM the MC-grown dust has a definite core mantle structure with
a more resilient core and a more easy destructible mantle. A theoretical
treatment of this grain structure would require to consider at least a two-phase
interstellar medium (cf. Tielens \cite{Tie98}; Inoue \cite{Ino03}), and not a
simple one-phase model as in our present calculation. 

The plan of our paper is as follows: In Sect.~\ref{SectChemEvol} our evolution
model for the solar neighbourhood of the Milky Way is introduced and some
results for the chemical evolution are discussed. In Sect.~\ref{SectDuEvol} the
model for the dust return by stars is explained. Section \ref{SectDestrGro}
discusses the dust destruction and growth processes in the ISM and 
Sect.~\ref{SectResDuEvol} presents the results for the evolution of the
interstellar dust. Some concluding remarks are given in Sect.~\ref{SectCR}.


\section{Chemical evolution}
\label{SectChemEvol}

To study the evolution of the dust content of our Galaxy we develop a standard
open model of galactic chemical evolution. In this model the Milky Way is formed 
by slow infall of primordial gas from the halo or intergalactic space. Merging with
other galaxies seems not to have played a mayor role during most of the lifetime
of the Milky Way, except for the very first evolutionary phase for which stellar
dynamics (cf. Helmi et al \cite{Hel06}) and elemental abundances (Reddy et al. 
\cite{Red06}; Ram\'{\i}rez et al. \cite{Ram07}) indicate that there were major
merging events. Merging is not considered in the model. In the one-zone
approximation we neglect radial motions in the galactic disk (but cf. Vorobyov
\& Shchekinov \cite{Vor06}) and consider its  evolution in a set of independent
rings. 

We solve numerically a classical set of non-linear integro-differential
equations for the chemical evolution of the Milky Way following a mathematical
formulation similar to Dwek (\cite{Dwe98}), who first extended the standard
system of equations for the chemical evolution of the galactic disk (cf.
Matteucci \cite{Mat03}) to include the evolution of the dust component of the
Galaxy. The basic set of equations for the surface densities of gas, stars,
nuclei, etc. is not repeated here but can be found in the papers cited before.
We specify in the following only our choices for some important input quantities
for the model calculations. 

\subsection{Basic model parameters}

\subsubsection{Infall}

First models of chemical evolution of the Galaxy were simple ``closed-box" 
models, in which the Galaxy's mass is already fixed at the initial instant of
evolution. However, these models failed to  reproduce the metallicity
distribution of metal poor stars, one of the most important observational
constraints on chemical evolution modeling, a problem being known as G-dwarf
problem. Open models assume formation of the Galaxy by accretion of primordial
or metal poor gas from extragalactic sources to solve the G-dwarf problem, as
was first suggested by Chiosi (\cite{Chi80}), and later discussed by Pagel
(\cite{Pag97}). In open models the total surface density of the disk changes by
the accretion of gas, outflows, and radial motions within the disk. For the
Milky Way outflows can be neglected due to the strong gravitational potential,
and radial motions are neglected in the one-zone approximation. In our model the
infall rate entirely defines the evolution of the total surface density.

Several scenarios for gas accretion have been proposed by different authors,
suggesting different rates and sequences of formation of the galactic
components, see  Matteucci (\cite{Mat03}) for details. Models assuming an
exponentially decreasing infall of the gas are most successful in reproducing
the G-dwarf distribution. Following Chiappini et al. (\cite{Chi97}) we adopt a
two-infall exponentially decreasing model that assumes two subsequent episodes
of Galaxy formation. Initially, the halo and thick disk are formed during a
short period of about $\tau_{\rm H}\approx1$~Gyr, then the thin disk is formed
by accretion of material on a much longer timescale of $\tau_{\rm D}\approx
7$~Gyr at Solar galactocentric radius (here $r_{\sun}=8.5$~\rm kpc).
The accretion rate is given in this model by the expression:
\begin{eqnarray}
{{\rm d}\,\Sigma_i(r,t)_{\rm inf}\over{\rm d}\,t}&=& \nonumber
\\
&&\hskip -1.8cm
\cases{(X_i)_{\rm inf}A(r)e^{-t/\tau_{\rm H}}
& for $t<t_{\rm thin}$\cr
(X_i)_{\rm inf}A(r)e^{-t/\tau_{\rm H}}
+(X_i)'_{\rm inf}B(r)e^{-(t-t_{\rm thin})/\tau_{\rm D}} &
for $t>t_{\rm thin}$}\,,
\end{eqnarray}
where $t_{\rm thin}=1$~Gyr is the time of onset of accretion onto the thin disk.
The formation of the disk is assumed to start $t_{\rm G}=13$ Gyrs ago.

The coefficients $A(r)$ and $B(r)$ are derived such as to reproduce the present
day density distribution of the disk. At the solar cycle one has
\begin{eqnarray}
A(r_{\sun})&=&{\Sigma_{\rm H}(r_{\sun},t_{\rm G})\over
\tau_{\rm H}\left(1-{\rm e}^{-t_{\rm G}/\tau_{\rm H}}\right)}\\
B(r_{\sun})&=&{\Sigma_{\rm tot}(r_{\sun},t_{\rm G})-\Sigma_{\rm H}
(r_{\sun},t_{\rm G})\over
\tau_{\rm D}\left(1-{\rm e}^{-(t_{\rm G}-t_{\rm thin})/\tau_{\rm D}}\right)}\,.
\end{eqnarray}
For details we refer to Chiappini et al. (\cite{Chi97}), Alibes et al.
(\cite{Ali01}). The value $\Sigma_{\rm tot}(r_{\sun},t_{\rm G})$ for the current
total density of the disk is taken to be $56\, \rm M_{\sun}\,pc^{-2}$ according
to Holmberg \& Flynn (\cite{Hol04}). For the contribution of the thick disk to
the total surface density we choose $\Sigma_{\rm H}=10\,\rm M_{\sun}\,pc^{-2}$.
$(X_i)_{inf}$ and $(X_i)'_{inf}$ denote element abundances of infalling gas,
which we assume to be primordial.

\subsubsection{Stellar birthrate}

Observations of global star formation rate in spiral galaxies suggest a
Schmidt-law type of dependence of the stellar birthrate on some power of the
total gas surface density $B\propto\Sigma_{\rm gas}^n$ with  $n~\approx1.5$
(Kennicutt \cite{Ken98}). An additional dependence of the star formation rate on
the total surface density $\Sigma_{\rm tot}(r,t)$ was suggested in self
regulating star formation theory (Talbot \& Arnett \cite{Tal75}). Later Dopita
\& Ryder (\cite{Dop94}) confirmed this by observations and suggested an empirical
law of star formation $B(r,t) \propto \Sigma_{\rm tot}^n \Sigma_{\rm g}^m$ with
$m=5/3$ and $n=1/3$ giving the best fit for the observed relationship between
the stellar brightness  and the surface brightness in H$_\alpha$ in galactic
disks. We adopt these values for the powers in the star formation law and
choose analogous to Alibes et al. (\cite{Ali01}) the following form of the star
formation rate: 
\begin{eqnarray}
B(r,t)=\nu \frac{\Sigma_{\rm tot}(r,t)^n \Sigma_{\rm g}(r,t)^m} {\Sigma_{\rm tot}
(r_{\sun},t)^{n+m-1}}\,.
\end{eqnarray}
We take into account a star formation threshold, i.e. a minimum surface density
$\Sigma_{\rm g}$ required for star formation, which is set to $7~\rm M_{\sun}\,
pc^{-2}$ (Kennicutt \cite{Ken98}). In the numerical calculation the transition
from zero to the threshold value is smoothed in order to avoid the unphysical
numerical oscillations in the solution for the surface density $\Sigma_{\rm g}$
close to the threshold that are produced otherwise by some integration methods.
The constant $\nu$ is fitted such that the model fits the present day star
formation rate in the solar neighbourhood. We take $\nu = 1.3~\rm Gyr^{-1}$.

\subsubsection{Stellar lifetimes}

We refuse the instantaneous recycling approximation, i.e. the assumption that
massive stars die immediately after their birth and return metals to the ISM,
and consider stellar lifetimes as a function of stellar mass and metallicity
using an analytical approximation given by Reiteri et al. (\cite{Rei96}). The
formula of Reiteri et al. (\cite{Rei96}) is a good fit for the stellar lifetimes
computed by the Padova group (Alongi et al. \cite{Alo93}; Bressan et al. 
\cite{Bre93}, Bertelli et al. \cite{Ber94}) in the metallicity range $7\cdot
10^{-5}<Z<3\cdot 10^{-2}$ and for initial masses  between 0.6 and 120~M$_{\sun}$.

\subsubsection{Initial mass function}
\label{SectIMF}

The initial mass function (IMF) $\Phi(M)$ gives the distribution of stellar
masses born in a star formation event. The IMF is one of the most important
components of the galactic chemical evolution model as it establishes the
frequency of low and high mass stars, and thus their relative role in the
chemical evolution. The IMF is commonly assumed to be constant in space and
time and is usually approximated by a power law $\Phi(M) = A M^{-\alpha}$, where
$A$ is a constant derived from a normalization of the IMF to unity in the
considered mass interval $\{M_{l}, M_{u}\}$. The first such an IMF for Solar
neighbourhood stars was proposed by Salpeter (\cite{Sal55}) with a power law
index $\alpha=2.35$, which is still widely used in chemical evolution models.
Later studies showed that the IMF is significantly flatter below 0.5~M$_{\sun}$
(e.g. Miller \& Scalo \cite{Mil79}; Kroupa \cite{Kro93}). For a discussion of
the concept of an IMF see Scalo (\cite{Sca05}) and Elmegreen \& Scalo 
(\cite{Elm06}). 

In the present paper we adopt a frequency distribution of stellar masses at
star formation consisting of four separate power-law type distributions in
four separate intervals of initial masses proposed by Kroupa (\cite{Kro02}):
\begin{equation}
\Phi(M) = A \left\{
\begin{array}{ll}
C_1 M^{-0.3}, & 0.01\le M/M_{\sun}< 0.08\\
C_2 M^{-1.3}, & 0.08\le M/M_{\sun} < 0.5\\
C_3 M^{-2.3}, & 0.5\le M/M_{\sun} < 1.0\\
C_4 M^{-2.7}, & 1.0\le M/M_{\sun} < 100,\\
\end{array}
\right.
\label{DefIMF}
\end{equation}
where the coefficients $C_1=2.0158$, $C_2=0.1612$, $C_3=C_4=0.0806$ are derived
from the normalization procedure. Masses are in solar masses. During the
calculation we found, that one obtains better model fits if for high mass stars
the exponent is changed to 2.55. Such a somewhat flatter power law ($\Phi\propto
M^{-2.6}$) is, for instance, observed for massive stars in the Orion nebula
(Preibisch et al. \cite{Pre02}).

The average stellar mass is then given by the integration over the full range of 
stellar masses:
\begin{equation}
M_{\rm av} = \int_{0.01}^{100}{M \Phi(M)\, dM}
\label{AverMass}
\end{equation}

\subsubsection{Nucleosynthesis Prescriptions}
\label{SectNuclSynth}

Nucleosynthesis prescriptions and stellar yields are another important ingredient
of chemical evolution modeling. Usually stellar yields are divided into three
main categories according to stellar masses, and result from extensive stellar
evolution calculations.

The single low and intermediate mass stars from the mass range 0.8-8 M$_{\sun}$
contribute to the enrichment of the Milky Way with heavy elements due to 
excessive mass-loss during the final stage of their AGB evolution. We adopt the
yields for H, $^4$He, $^{12}$C, $^{13}$C, $^{14}$N, and $^{16}$O from van den
Hoek \& Groenewegen (\cite{Hoe97}), tabulated  for the range 0.8 -- 8 M$_{\sun}$
of initial masses and for metallicities from $10^{-3}$ to $4 \cdot 10^{-2}$.
For $^{23}$Na, $^{24}$Mg, $^{25}$Mg, $^{26}$Mg, $^{26}$Al, and $^{27}$Al the
yields of Karakas et al. (\cite{Kar03}) for the mass-range 1.0 -- 6.5 
M$_{\sun}$ and range of metallicities $Z=0.004,\ 0.008,\ 0.02$ are used. Outside
of the range of tables the data are extrapolated.

Rates of SN Ia explosions are calculated in the approximation of Matteucci \&
Greggio (\cite{Mat86}), based on the classical scenario of deflagration in C-O
White Dwarfs in binary systems (Whelan \& Iben \cite{Whe73}), with a
modification recently proposed by Hachisu et al. (\cite{Hac96, Hac99}) that
accounts for the important effect of metallicity on mass transfer in binaries to
a compact object. In this scenario the accreting White Dwarf develops an
optically thick wind. If ${\rm [Fe/H]}<-1$ the wind is too weak for a SN Ia to
occur; only for higher metallicities one has a contribution from SN Ia
explosions to the heavy element production. The SN Ia yields are taken from
Iwamoto et al. (\cite{Iwa99}). The parameter determining the frequency of events
is fitted such that the iron abundance of the Solar System is reproduced; a
value of $\beta=2\times10^{-2}$ was found to give the best result.

The problem of yields from massive stars is complicated by the necessity to
model supernova explosions, many details of which are still unknown. We adopt
the recent nucleosynthesis prescriptions by Nomoto et al. (\cite{Nom06}) which
presented SN~II nucleosynthesis yields as a functions of stellar masses (from
the range of 11-50~$\rm M_{\sun}$), metallicity, and explosion energy. Two
distinct new classes of massive SNe are taken into account in this work: very
energetic Hypernovae, and very faint and low energy SNe. Nucleosynthesis in
hypernovae can explain the observed trends of (Zn, Co, V, Ti, Mn, Cr)/Fe in
extremely metal poor stars (Nomoto et al. \cite{Nom06}; Kobayashi et al.
\cite{Kob06}). For comparison purposes we implemented also the yields of
Woosley \& Weaver (\cite{Woo95}) for stars with masses from the range of
11-40~$\rm M_{\sun}$, that are most commonly used in chemical evolution
calculations.

For massive stars with $M>40\,{\rm M}_{\sun}$ the mass returned by the stars up
to the end of carbon burning is taken from the models of Schaller et al.
(\cite{Sch93}), Schaerer et al. (\cite{Sch93}), and Charbonnel et al. 
(\cite{Cha93}).  It is assumed that the remaining mass collapses into a Black
Hole. The mass-return of nuclei is determined from the models for all those
nuclei, for which  surface abundances are given in the tables. For all others we
assume that their abundance in the returned mass equals their initial  abundance.

It is assumed that the mass returned by stars is mixed with the general
interstellar medium on timescales much shorter than the timescale for conversion
into new stars, i.e., the composition of the interstellar medium is assumed to be
homogeneous at each instant. This is justified by the observed low scatter of
element abundances in the present ISM and of stellar element abundances in open
stellar clusters (see Scalo \& Elmegreen \cite{Sca04}, and references therein).

\begin{figure}[t]
\resizebox{\hsize}{!}{\includegraphics{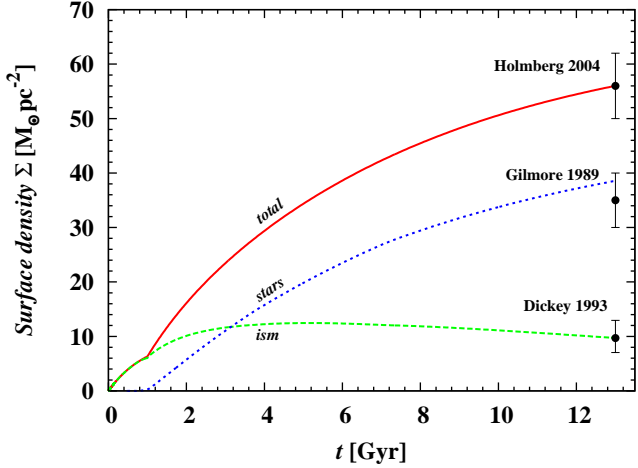}}

\caption{Time evolution of the total surface density $\Sigma_{\rm tot}$ (full
line), of the surface density of visible stars (dotted line), and of the
interstellar matter surface density $\Sigma_{\rm ISM}$ (dashed line) of the
galactic disk at the solar distance from the galactic centre, and observed values
for the present day total surface density in the solar neighbourhood (Holmberg
\& Flynn \cite{Hol04}), for the surface density of the stellar component
(Gilmore et al. \cite{Gil89}), and for the surface density of the interstellar
medium (Dickey \cite{Dic93}).
}
\label{GFigEvolDisk}
\end{figure}

\subsection{Evolution of some disk properties}

We now show some results of a numerical calculation of the galactic disk's 
chemical evolution at the solar cycle that are important for our problem. For
the model presented in the following the nucleosynthetic yields of massive stars
are taken from the tables of Nomoto et al. (\cite{Nom06}). 

Figure \ref{GFigEvolDisk} shows the evolution the total surface mass density
$\Sigma_{\rm tot}(r_{\sun},t)$ and that of the interstellar medium 
$\Sigma_{\rm ISM}(r_{\sun},t)$ for the galactic disk. In the model it is assumed
that the formation of the disk started 13\,Gyrs ago. Initially most of the
material in the disk was in gaseous interstellar matter, today and at the time
of Solar System formation most of the galactic matter is condensed into stars.
A minor fraction is locked up in stellar remnants (White Dwarfs, Neutron Stars,
Black Holes).

\begin{figure}[t]

\resizebox{\hsize}{!}{\includegraphics{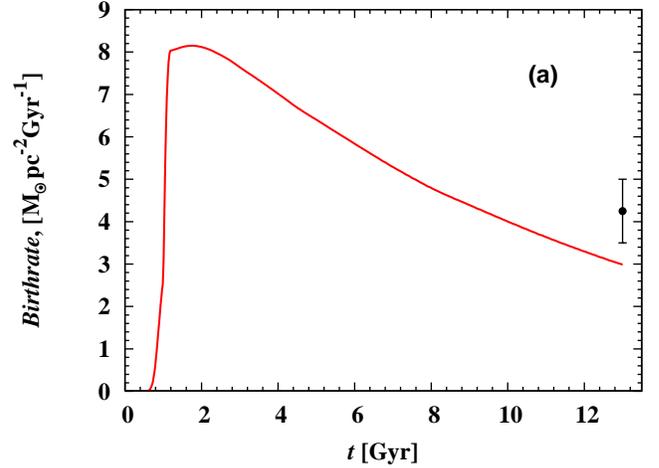}}

\resizebox{\hsize}{!}{\includegraphics{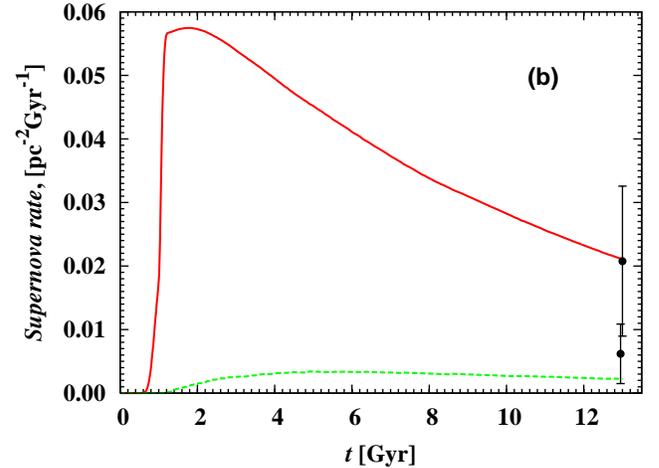}}

\caption{{\bf(a)} Evolution of the astration rate $B$ at the solar
galactocentric distance $r_{\sun}$. The errorbar shows the presently observed
stellar birthrate (Rana \cite{Ran91}). {\bf(b)} Evolution of the supernova type
II (full line) and type Ia (dashed line) rates at the solar galactocentric
distance $r_{\sun}$ and observed values at present time from Tammann
(\cite{Tam94}).
}
\label{FigBirthR}
\end{figure}

The evolution of the stellar birthrate is shown in Fig.~\ref{FigBirthR}a. Star
formation commences about 1 Gyr after the onset of matter infall since about 1
Gyr time is required in the two-infall model until the gas density at the
galactocentric distance of the sun increases to the threshold value for star
formation of $M_{\rm ISM}=7\,\rm M_{\sun}$ (Kennicutt \cite{Ken98}). The stellar
birthrate culminated about 2 Gyrs after the onset of star formation and since
then it gradually declines. Most of the stars born are low and intermediate
mass stars; the massive stars mass fraction of the newly born stars is only
6.5\% according to the initial mass function Eq.~(\ref{DefIMF}), but this small
fraction is responsible for nearly all of the heavy nuclei synthesized in the
Milky Way. 

Figure \ref{FigBirthR}b shows the evolution of the supernova rates at the solar
galactocentric distance $r_{\sun}$. Because of the short lifetime of massive
stars the supernova rates for type~II supernovae closely resembles the birthrate
of stars. Supernovae of type Ia appear with a delay of several Gyrs because (i)
their progenitors are long lived intermediate mass stars and (ii) the suppression
of supernova explosions in binaries at low metallicities proposed by Hachisu et
al. (\cite{Hac96,Hac99}). Since supernovae of type Ia are the main sources of
Fe, the iron abundance increases in the Milky Way only on a rather long
timescale.

\subsection{Chemical evolution of the disk}

The viability of galactic chemical evolution models is usually tested by
comparison with some standard observational constraints: G-dwarf metallicity
distribution, age-metallicity relation, Solar System abundances at the instant
of its formation $t_{\rm SSF}$, and evolution of element abundance ratios over
time.

\begin{figure}[t]
\resizebox{\hsize}{!}{\includegraphics{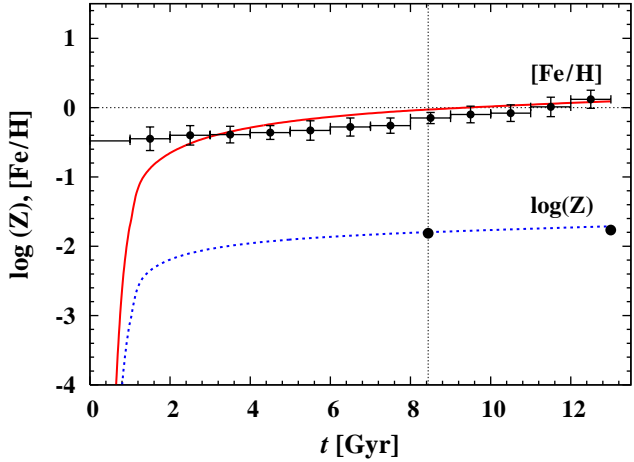}}

\caption{Evolution of metallicity $Z$ of the ISM at the solar galactocentric
distance $r_{\sun}$ and of the [Fe/H] abundance ratio. The error bars show the
observed age-metallicity relation from Rocha-Pinto (\cite{Roc00}). The thin
vertical line indicates the Solar System birth time, and the two filled circles
indicate the observed metallicity of the sun and of the present day ISM}
\label{FigEvolMet}
\end{figure}

\subsubsection{Evolution of metallicity}

Figure~\ref{FigEvolMet} shows the time evolution of the metallicity $Z$ and the
abundance ratio [Fe/H]\footnote{
The abundance ratio [X/Y] for two elements X and Y is defined as
\begin{displaymath}
\textrm{[X/Y]}=\log\left({\epsilon_{\rm X}\over\epsilon_{\rm Y}}\right)-
\log\left({\epsilon_{\rm X}\over\epsilon_{\rm Y}}\right)_{\sun}\,,
\end{displaymath}
where $\epsilon_{\rm X}$ is the element abundance of element X by number
relative to hydrogen.}
of the interstellar medium at the solar radius $r_{\sun}$.
The predicted evolution of the [Fe/H] ratio in the ISM is
compared with the age-metallicity relation of the solar neighbourhood of
late-type dwarfs from Rocha-Pinto (\cite{Roc00}). The thin black vertical line
shows the instant of Solar System formation $t_{\rm SSF}$, which we assume to be
4.56~Gyr ago. The filled circles indicate the observed Solar metallicity
$Z_{\sun}$ and the present day ISM metallicity. The model reproduces the
observational values quite well. 

All model calculations for the evolution of heavy element abundances with time
predict a well defined relation between metallicity and time-of-birth at a
certain location in the galactic disk like that shown in Fig.~\ref{FigEvolMet}.
Observationally determined ages obtained by comparing the position of a star in
the Hertzsprung-Russel diagram with evolutionary isochrones, and relating
spectroscopically determined metallicities of stars with such age
determinations, show a tremendous scattering of metallicities for a given age. It
has been concluded that this reflects (i) a true scattering of metallicities of
the matter out of which stars are formed at given galactocentric radius and
birthtime, and (ii) possibly a mixing of stars from different galactic zones by
radial diffusion (Edvardsson et al. \cite{Edv93}). Pont \& Eyer (\cite{Pon04})
have shown, however, that the tremendous scattering results most likely from the
difficulty of obtaining reliable stellar ages from evolutionary isochrones and
that any true internal scattering of metallicities at given age is probably less
than 0.15 dex. More careful analysis of age-metallicity relations based on such
improved methods (da Silva et al. \cite{daS06}) also seem to support a small
intrinsic scattering of metallicities at a given age. The age-metallicity
relation of Rocha-Pinto (\cite{Roc00}) is based on ages determined from
spectroscopic indicators of chromospheric activity, which cannot be considered
as particular accurate, but also this points to a small internal scattering of
metallicities. 

\begin{figure}[t]
\resizebox{\hsize}{!}{\includegraphics{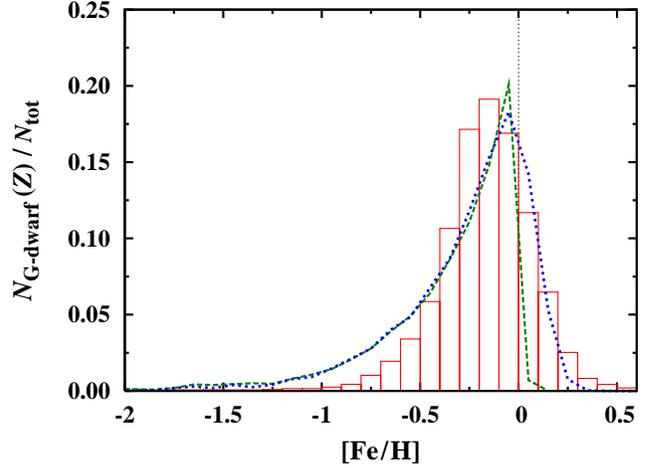}}

\caption{G-dwarf metallicity distribution in the solar vicinity predicted by the
model and the observed distribution as derived by Nordstr\"{o}m (\cite{Nor04}).
The thin dashed line shows the G-dwarf distribution from direct calculations,
while the thick dotted line is the result of a convolution with a Gaussian with
halfwidth 0.2~dex to account for the observational scatter.
}
\label{FigGdwarf}
\end{figure}

The age-metallicity relation is reasonably well reproduced for the last
about 10 Gyrs, but there is an increasing discrepancy for earlier times. This is
a general problem of all such evolution calculations and results most likely 
(i) from the assumption that the infalling material has primordial abundances
while in fact there is some pre-enrichment with heavy elements in this material,
and (ii) from unrealistically high stellar ages for many stars due to the rather
crude methods of age determination.

\subsubsection{Metallicity distribution of G dwarfs}

The G-dwarf metallicity distribution is one of the most important observational
tests, since lifetimes of G-dwarf stars are comparable with the Galactic age,
so that their metallicity distribution reflects past star formation history. We
compare the observed G-dwarf distribution from the most recent and most
complete compilation of Nordstr\"om et al. (\cite{Nor04})  with that
predicted by the model in Fig.~\ref{FigGdwarf}. The dotted line shows the direct
calculational results, while the dashed line displays a convolution of the model
results with a Gaussian with dispersion of 0.2~dex in order to simulate
observational errors in the metallicity determination and intrinsic cosmic
scatter in metal abundances. The errors of modern abundance determinations are
usually claimed to be 0.1 dex or even less. The true scatter of stellar
abundances for stars born at the same instant and location is difficult to
determine since for single stars neither their birthplace nor their birthtime is
accurately known. The small scatter of abundances between stars in open stellar
clusters indicate, however, that the intrinsic scatter seems to be very small
(see Scalo \& Elmegreen \cite{Sca04}, and references therein); we arbitrarily
assume a contribution of 0.1\,dex to the total scatter.

\begin{figure*}[t]

\hspace{0.10\hsize}
\resizebox{0.75\hsize}{!}{\includegraphics{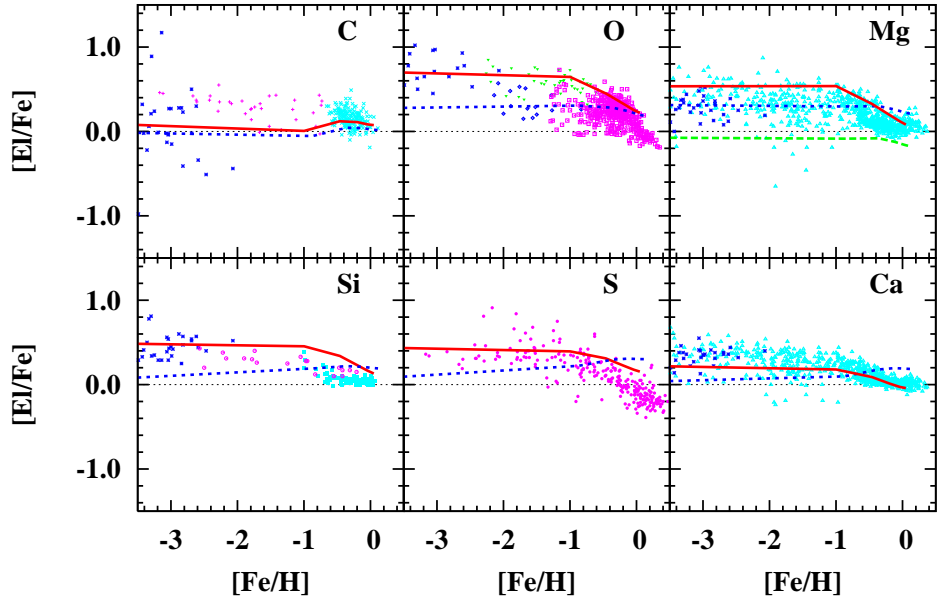}}

\caption{Comparison of the predicted abundance ratios of main dust forming
elements [El/Fe] with observations of stellar abundances. The solid and dashed
lines show model calculations with Nomoto (\cite{Nom06}) and Woosley \& Weaver
(\cite{Woo95}) SNII yields, respectively. We corrected  WW95 yields for Fe and
Mg to achieve better fits to observations. For illustrative purposes a model
calculation with uncorrected Mg yields from WW95 is shown with a thin dashed
line. The observed stellar element abundances for F and G stars from the solar
neighbourhood are shown with different symbols for each of the sources
(Akerman et al. \cite{Ake04}, Reddy et al. \cite{Red03}, Soubiran et al. 
\cite{Sou05}, Melendez et al. \cite{Mel02}, Jonsell et al. \cite{Jon05}, Venn
et al. \cite{Ven04}, Chen et al. \cite{Che00}, Gratton et al. \cite{Gra91},
Caffau et al. \cite{Caf05}, Cayrel et al. \cite{Cay04}).}
\label{FigElRat}
\end{figure*}

The calculated metallicity distribution reproduces the general trends of the
observed distribution, but it does not agree particular well. Significant
deviations are seen for low and high metallicities. After convolution the 
discrepancies at the higher metallicity end disappear almost completely. The
theoretical result then reasonably well agrees with the observations. For low
metallicities the discrepancies persist and indicate that our model assumptions
are likely not realistic for the earliest evolutionary phase. Since for the main
application of our model this phase is not important, we did not try to improve
the model in this respect.

\subsubsection{Evolution of abundances of individual elements}

Abundance ratios [X/Fe] of elements X and their variation with time reflect the
synthesis of heavy elements during galactic evolution. The reproduction of these
variations by the model is one of the most important tests for the reliability
of the model. For comparing the variation of [X/Fe] with observed variations of
stellar abundances, stellar ages would be required which, however, are unknown
or of low accuracy. One prefers to compare instead the variation of the
abundance ratios [X/Fe] with the abundance ratio [Fe/H], since [Fe/H] is also
determined from stellar atmosphere analysis and varies, at least for the Milky
Way, monotonously with age of the galactic disk (cf. Fig.~\ref{FigEvolMet}),
i.e., can be taken as a measure of stellar age. We have calculated in our model
the evolution of 63 isotopes using nucleosynthesis prescription of Nomoto et al.
(\cite{Nom06}) and, for comparison, that of Woosley \& Weaver (\cite{Woo95}).
Results are presented in Fig.~\ref{FigElRat} for the elements related to dust
formation. We concentrate here on these elements, since in the present work we
are mainly concerned with problems related to interstellar dust evolution.

The figure shows as thick lines the model results if SN II yields from Nomoto et
al. (\cite{Nom06}) are used, and as thick dashed lines the corresponding results
if yields from  Woosley \& Weaver (\cite{Woo95}) are used. The various dots,
crosses etc. show results of stellar abundance analysis for G stars from the
solar neighbourhood; the sources of data are given in the figure caption. These
data show a considerable scatter because of the errors of abundance
determinations and possibly some small intrinsic scatter of element abundances
of stars of comparable age. Nevertheless there are clear observable correlations
between  the abundance ratios [X/Fe] and [Fe/H]. For the elements shown the new
results of Nomoto et al. give better agreement between the calculated abundance
evolution and the observed correlations of [X/Fe] with [Fe/H] than the older
Woosley \& Weaver results, for other elements, however, there are some
discrepancies with observations.

With the yields of Woosley \& Weaver (\cite{Woo95}) there are some substantial
problems. First, the iron yields of Woosley \& Weaver are too high, as
already found in Timmes et al. (\cite{Tim95}), and we follow their
recommendation to reduce the Fe yield.  Second, there is another severe problem
with the Woosley \& Weaver results for magnesium. The calculated abundances
based on the original yields are shown in Fig.~\ref{FigElRat} with a thin dashed
line. These abundances are definitely too low, a problem which  is known since
long (e.g. Goswami \& Prantzos \cite{Gos00}; Francois et al. \cite{Fra04}). A
comparison with the observed evolution of stellar magnesium abundances with
metallicity shows that the shape of the [Mg/H]-[Fe/H]-relation is reasonably well
reproduced by the model, except that the absolute values of [Mg/H] are
systematically too low by a factor of 2.5. We have increased therefore the Mg
yields of Woosley \& Weaver (\cite{Woo95}) by this factor in order to reproduce
the Mg abundance of the Solar System. The resulting variation of [Mg/H] with
[Fe/H] is shown in the figure with a dashed line, which reproduces the
observations much better. Such a correction would be necessary for the purpose
of calculating dust abundances, since reliable results for dust condensation
require that the abundance ratios Si/Mg and Fe/Si of the main dust-forming
elements agree with the observed abundance ratios in the Milky Way. Otherwise
one would get a deviating dust mixture.

Since the yields of Nomoto et al. (\cite{Nom06}) gives results for the abundance
evolution of the main dust forming refractory elements much closer to
observations than the Woosley \& Weaver (\cite{Woo95}) yields, and since they do
not require for this to introduce some ad hoc scalings, we prefer to use the
Nomoto et al. (\cite{Nom06}) yields for the model calculations.

\begin{figure}

\centerline{
\includegraphics[width=.8\hsize]{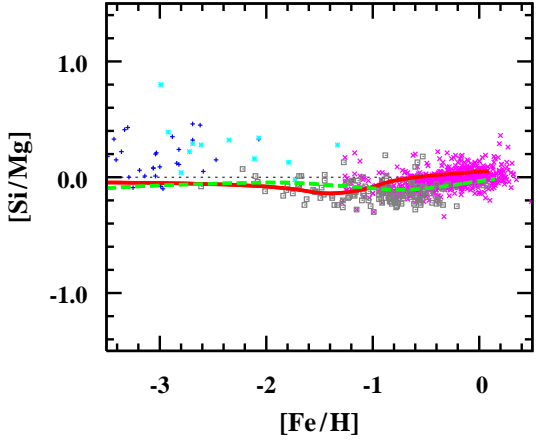}
}

\caption{Abundance ratio Si/Mg of the major silicate dust forming elements.
The full line corresponds to a model using SN yields of Nomoto et al. 
(\cite{Nom06}), the dashed line to a model using SN yields from Woosley \& Weaver
(\cite{Woo95}). For the latter the Mg abundance is scaled such that it
reproduces the solar Mg abundance at [Fe/H]=0.}
\label{FigMgSiRatio}
\end{figure}

Figure \ref{FigMgSiRatio} compares the (adjusted) calculated abundance ratio
Si/Mg with observed abundance ratios in the atmospheres of nearby F and G stars
and their correlation with metallicity [Fe/H]. The model results are close to
the observed values. The Si/Mg ratio determines the nature of the silicate dust
that can be formed if Mg and Si both are completely condensed into dust. Oxygen
is in any case abundant enough for formation of any kind of Mg-Si-compound. For
a ratio of Si/Mg = 1 one can form enstatite (MgSiO$_3$), for Si/Mg = 2 one can
form forsterite (Mg$_2$SiO$_4$). For values in between a mixture of both can
be formed; here part of the Mg can be replaced by Fe and one rather forms a
mixture of magnesium-iron-silicates. The stellar data show that the Si/Mg ratio
is close to unity for metallicities $\rm [Fe/H]>-1$, and therefore one observes
the formation of Mg-Fe-silicates in space.

\begin{table}

\caption{Solar system element abundances $a$ and the standard error of the
abundance determination $\sigma$. Abundances for the solar photosphere are from
the compilation of Asplund et al. (\cite{Asp05}), except for He where the
recommended value for the early sun from Grevesse \& Sauval (\cite{Gre98}) is
given. Abundances for meteorites are from the compilation of Palme \& Jones
(\cite{Pal02}). For C, N and O also abundances from Holweger (\cite{Hol01})
are given. For Ne and Ar see text. The last column indicates if solar
photospheric abundances from Asplund et al. (S), or from Holweger (H), or
meteoritic (M) abundances are preferred for comparison with model calculation
results; a small letter indicates that the element is not used for a comparison
in the present work.
}

\begin{tabular*}{\hsize}{rl@{\hspace{11mm}}cc@{\hspace{11mm}}cc@{\hspace{1cm}}c}
\hline\hline
\noalign{\smallskip}
$Z$ & Elem. & \multicolumn{2}{c}{Sun} & \multicolumn{2}{c}{Meteorites}& used \\
    &       & $a$ & $\sigma$   & $a$ & $\sigma$ &  \\
\noalign{\smallskip}
\hline
\\
 1 & H  & 12.00 &   -- &      &      & S \\
 2 & He & 10.99 & 0.02 & 1.92 &      & S \\
 3 & Li &  1.05 & 0.10 & 3.30 & 0.04 & m \\
 4 & Be &  1.38 & 0.09 & 1.41 & 0.04 & m \\
 5 & B  &  2.70 & 0.20 & 2.77 & 0.04 & m \\
 6 & C  &  8.39 & 0.05 & 7.39 & 0.04 &   \\
   &    &  8.59 & 0.11 &      &      & H \\
 7 & N  &  7.78 & 0.06 & 6.32 & 0.04 &   \\
   &    &  7.93 & 0.11 &      &      & H \\
 8 & O  &  8.66 & 0.05 & 8.43 & 0.04 &   \\
   &    &  8.74 &      &      &      & H \\     
 9 & F  &  4.56 & 0.30 & 4.45 & 0.06 & m \\
10 & Ne &  8.08 & 0.07 &      &      & s \\
11 & Na &  6.17 & 0.04 & 6.30 & 0.02 & M \\
12 & Mg &  7.53 & 0.09 & 7.56 & 0.01 & M \\
13 & Al &  6.37 & 0.06 & 6.46 & 0.01 & M \\
14 & Si &  7.51 & 0.04 & 7.55 & 0.01 & M \\
15 & P  &  5.36 & 0.04 & 5.44 & 0.04 & S \\
16 & S  &  7.14 & 0.05 & 7.19 & 0.04 & M \\
17 & Cl &  5.50 & 0.30 & 5.26 & 0.06 & s \\
18 & Ar &  6.70 & 0.06 &      &      & s \\
19 & K  &  5.08 & 0.07 & 5.11 & 0.02 & M \\
20 & Ca &  6.31 & 0.04 & 6.33 & 0.01 & M \\
22 & Ti &  4.90 & 0.06 & 4.95 & 0.04 & M \\
23 & V  &  4.00 & 0.02 & 3.99 & 0.02 & M \\
24 & Cr &  5.64 & 0.10 & 5.67 & 0.01 & M \\
25 & Mn &  5.39 & 0.03 & 5.51 & 0.01 & M \\
26 & Fe &  7.45 & 0.05 & 7.49 & 0.01 & M \\
27 & Co &  4.92 & 0.08 & 4.90 & 0.01 & M \\
28 & Ni &  6.23 & 0.04 & 6.23 & 0.02 & M \\
29 & Cu &  4.21 & 0.04 & 4.28 & 0.04 & M \\
30 & Zn &  4.60 & 0.03 & 4.66 & 0.04 & M \\
\noalign{\smallskip}
\hline
\end{tabular*}

\label{TabAbuSolSy}
\end{table}

\subsubsection{Solar System abundances}

The model should also reproduce the element abundances of the Solar System since
they reflect the ISM composition at $r_{\sun}$ at the instant $t_{\rm SSF}=
4.56$\,Gyr ago, when the Sun was formed. Table~\ref{TabAbuSolSy} shows element
abundances in the Solar System in the frequently used logarithmic scale
($\epsilon$ is the abundance of an element relative to~H by number)
\begin{equation}
a=\log\epsilon+12
\end{equation}
for the elements from H to Zn that can be compared to the results of the model
calculation if we use supernova yields from Woosley \& Weaver (\cite{Woo95}) or
Nomoto et al. (\cite{Nom06}), since the tables cover only this range of
elements. Element abundances for the Solar System are available from either
spectroscopic abundance determinations from the solar photosphere or from
laboratory analysis of primitive meteorites. For the solar photosphere the
table gives data from the compilation of Asplund, Grevesse \& Sauval 
(\cite{Asp05}). For meteoritic abundances the table shows data from the
compilation of Palme \& Jones (\cite{Pal02}). 

The tabular values for the photosphere consider the recent significant
downward revision of the abundances of O, C, and N by Allende Prieto, Lambert
\& Asplund (\cite{All01,All02}) compared to the previous compilations of
Grevesse \& Sauval (\cite{Gre98}) and Anders \& Grevesse (\cite{And89}). The
table also gives the abundances for C, N, and O derived by Holweger
(\cite{Hol01}), who also found a reduction of the solar abundances for these
elements to be necessary, but not as much reduced as in the papers by Allende
Prieto et al. The abundances of Allende Prieto et al. pose serious problems to
solar helioseismology (Delahaye \& Pinsonneault \cite{Del06}; Basu et al. 
\cite{Bas07}) while the abundances of Grevesse \& Sauval  (\cite{Gre98}) give
good fits to observations. The incompatibility of the new C, N, O abundances with
helioseismological results should be taken serious and the abundance reductions
following from using numerically calculated models for the solar convective
flows (Asplund et al. \cite{Asp00}) to determine spectral line profiles seem to
result in unrealistically small abundances. Also comparison with abundances in
nearby G stars seem to indicate this (see section \ref{SectAbuFG}). The spatial
resolution of their flow calculations of about 50 km (Asplund et al. 
\cite{Asp00}) compared to a pressure scale-height of the solar photosphere of
about 125 km probably is insufficient and does not allow to account for velocity
fluctuations on length scales small compared to the mean free path lengths of
photons and therefore produces incorrect equivalent widths.

For the solar photosphere no reliable abundances of the noble gases can be
determined. For He a photospheric abundance is given in the table which is the
value recommended by  Grevesse \& Sauval (\cite{Gre98}) to be taken as the value
of the He abundance of the  early sun; the He abundance of the present sun is
lower due to segregation effects and cannot be used for a comparison. The
abundances for the other noble gases given in the table are determined 
from the Ne/Mg and Ar/Mg abundance ratios determined from coronal lines as given
by Feldman \& Widing (\cite{Fel03}). It is not sure that they really correspond
to the initial solar abundances. 

For meteorites the abundance of H and of the noble gases do not reflect their
abundance in the material out of which the parent bodies of the meteorites
formed since these elements are not incorporated into the bodies of the early
Solar System. Therefore no data for meteoritic abundances are given in the table
for these elements. 

\begin{figure}[t]
\resizebox{\hsize}{!}{\includegraphics{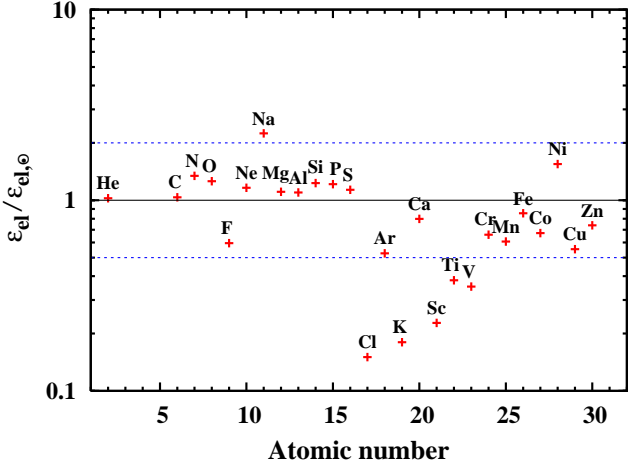}}

\caption{Calculated element abundances relatively to solar abundances at the
instant of Solar System formation (data according to Table \ref{TabAbuSolSy}).
Thin dotted lines show an as much as twice deviation from observed values
}
\label{FigPla}
\end{figure}

For meteorites the abundances of the volatile elements C, N, and O also are not
representative for the abundances in the early Solar System since these elements
are not (N) or only to a small fraction (C, O) condensed into solids and
incorporated into the parent bodies of the meteorites. Correspondingly, the
meteoritic abundances of C, N, and O given in the table are much lower than the
photospheric abundances. For these elements the solar photospheric abundances
have to be used for comparison purposes.
A number of elements are highly volatile (cf., e.g., Palme  \& Jones
\cite{Pal02}) and it is doubtful that these elements are completely condensed in
the parent bodies of the meteorites. Besides H, the noble gases, and C, N, and
O, these are the elements Cl, Br, I, In, Cs, Hg, Tl, Pb, Bi, from which Cl is
one of the elements in the table. For comparison purposes one should therefore
use in principle the Cl abundance from the photosphere, but since the abundance
determination of Cl for the solar photosphere is rather inaccurate, Cl is
presently not suited for comparison purposes.

For the remaining elements, the meteoritic and the solar photospheric abundances
well agree, except for a number of heavier elements not contained in our table.
For comparison with the results of the chemical evolution calculation, we usually
preferred the more accurate meteoritic abundance, whereas the solar photospheric was used when both methods were of only moderate accuracy (as specified in last
column of Table \ref{TabAbuSolSy})

\begin{table*}

\caption{Average abundances $a$ of F and G stars with solar metallicity 
(\,$\left\vert\Delta\rm[Fe/H]\right\vert<0.05$\,) and of young stars (age 
$\le 1\,$Gyr) from the solar neighbourhood. $\sigma_{\rm abd}$ is the
accuracy of the abundance determinations from stellar spectra, $\sigma_{*}$ is
the scattering of the stellar abundances around the mean. $Z$ is the metallicity
calculated from these abundances
}

\begin{tabular*}{\hsize}{@{\extracolsep\fill}rll|cccccc|ccc|ccc}
\hline\hline
 & &\rule[-3mm]{0mm}{8mm} & 
 \multicolumn{6}{c|}{Nearby F \& G stars}&
 \multicolumn{3}{c|}{B dwarfs} &
 \multicolumn{3}{c}{Orion nebula}\\
 & & & \multicolumn{3}{c}{solar met.} & \multicolumn{2}{c}{age $<1$ Gyr} &
& & & & & & \\
Z \rule[-2mm]{0mm}{8mm}&\multicolumn{2}{c|}{Element} & $a$ & $\sigma_{\rm abd}$ & $\sigma_{*}$ 
& $a$ & $\sigma_{*}$ & Source & $a$ & $\sigma_{*}$ & Source &
 $a$ & $\sigma_{\rm abd}$ & Source  \\
\hline
&&&&&&&&&&&&&&\\
 2 &      & He &       &      &      &      &      &   & 11.02 & 0.05 & 3 &
   10.99& 0.01 & 10 \\
 6 &\quad & C  &  8.37 & 0.06 & 0.11 & 8.39 & 0.11 & 2 & 8.32  & 0.10 & 4 &
   8.52 & 0.05 & 10 \\
 7 &      & N  &       &      &      &      &      &   & 7.73  & 0.28 & 5 &
   7.73 & 0.09 & 10 \\
 8 &      & O  &  8.75 &      & 0.07 & 8.77 & 0.13 & 1 & 8.63  & 0.18 & 4 &
   8.73 & 0.09 & 10 \\
10 &      & Ne &       &      &      &      &      &   & 8.11  & 0.04 & 6 &
   8.05 & 0.07 & 10 \\
11 &      & Na &  6.30 & 0.03 & 0.16 & 6.27 & 0.10 & 1 & & & & & & \\
12 &      & Mg &  7.63 & 0.06 & 0.32 & 7.64 & 0.21 & 1 & 7.59  & 0.15 & 7 &
     & & \\
13 &      & Al &  6.52 & 0.05 & 0.24 & 6.54 & 0.22 & 1 & 6.24  & 0.14 & 5 &
     & & \\
14 &      & Si &  7.60 & 0.05 & 0.28 & 7.61 & 0.23 & 1 & 7.50  & 0.21 & 5 &
     & & \\
16 &      & S  &  7.17 & 0.16 &      & 7.29 & 0.10 &   & 7.22  & 0.10 & 4 &
   7.22 & 0.04 & 10 \\
18 &      & Ar &       &      &      &      &      &   & 6.48  & 0.04 & 8 &
   6.62 & 0.05 & 10 \\
20 &      & Ca &  6.42 & 0.03 & 0.37 & 6.48 & 0.39 & 1 & & & & & & \\
22 &      & Ti &  4.92 & 0.03 & 0.11 & 4.94 & 0.14 & 1 & & & & & & \\
24 &      & Cr &  5.66 & 0.02 & 0.13 & 5.73 & 0.28 & 1 & & & & & & \\
26 &      & Fe &  7.55 & 0.06 & 0.12 & 7.61 & 0.25 & 1 & 7.46  & 0.08 & 9 &
     & & \\
28 &      & Ni &  6.22 & 0.02 & 0.09 & 6.25 & 0.07 & 1 & & & & & & \\
30 &      & Zn &  4.53 & 0.06 & 0.27 & 4.54 & 0.13 & 1 & & & & & & \\
&&&&&&&&&&&&&&\\
\rule[-2mm]{0mm}{4mm}   &      & $Z$& 0.0194&      &      & 0.0203 & & & 
  0.168 & & & & & \\
\hline
\end{tabular*}

\medskip\noindent Sources:
(1) Bensby et al. (\cite{Ben05}),
(2) Bensby \& Feltzing (\cite{Ben06}),
(3) Lyubimkov et al. (\cite{Lyu04}),
(4) Daflon \& Cunha (\cite{Daf04}),
(5) Rolleston et al. (\cite{Rol00}),
(6) Cunha et al. (\cite{Cun06}),
(7) Lyubimkov et al. (\cite{Lyu05}), 
(8) Holmgren et al. (\cite{Hol90})
(9) Cunha \& Lambert (\cite{Cun94}),
(10) Esteban et al. (\cite{Est04}),
\label{TabElAbBen05}
\end{table*}

Additionally, the elements Li, Be, B, F are excluded from the comparison, since
their production mechanisms are not implemented in the model program.

In Fig.~\ref{FigPla} we present the predicted element abundances of the ISM
relative to Solar System abundances at the instant of Solar System formation
at $r=r_{\sun}$. Thin horizontal lines indicate a deviation by a factor of
two upward or downward from Solar System abundances. As can be seen, the model
fits the observed abundances with good accuracy. Most calculated element
abundances reproduce the Solar System element abundances within a factor of 
about two, many elements even much better. The somewhat worse results for
Cl, K, and Sc have also be found by Kobayashi et al. (\cite{Kob06}); the rather bad
results for these elements are of no importance for our model, since they are
not one of the main dust forming elements.

\begin{figure}[t]
\centerline{
\resizebox{\hsize}{!}{\includegraphics{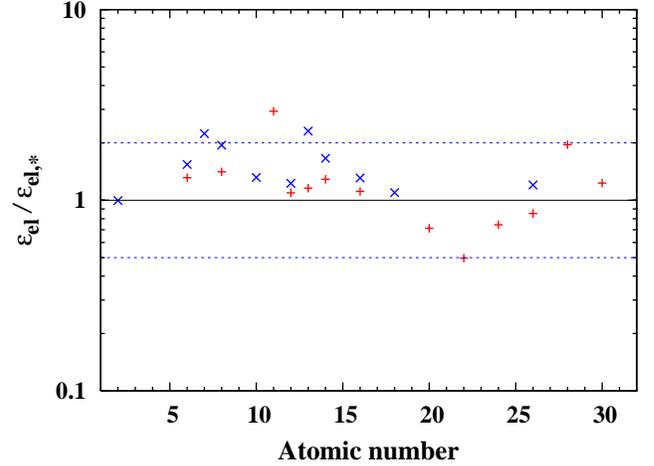}}
}
\caption{Calculated element abundances relative to abundances of F \& G stars
from the solar vicinity with with ages less than 1 Gyr (plusses), and of B
stars from the range $r_{\sun}\pm2\, \rm kpc$ of galactocentric distances
(crosses) (data according to Table \ref{TabElAbBen05}). Thin dotted lines show
an as much as twice deviation from observed values
}
\label{FigAbuFGB}
\end{figure}

\subsubsection{Abundances of F \& G stars}
\label{SectAbuFG}

Abundances of the ISM cannot be measured directly because in the ISM the
refractory elements are condensed into dust (cf. Savage \& Sembach 
\cite{Sav96}). One possibility to determine indirectly total element abundances
in the ISM is to determine atmospheric abundances from some kind of `young'
stars which have not changed their surface abundances since they formed from
interstellar matter. Best suited for this purpose are probably F and G main
sequence stars from the galactic neighbourhood, that show the kinematics of thin
disk stars and high metallicities, or, if stellar ages have been determined, are
of an age of no more than a few Gyrs. For such stars one can assume that they
formed from interstellar material with essentially the same properties as the
present day ISM of the galactic neighbourhood. Bensby et al. (\cite{Ben05}) and
Bensby \& Feltzing (\cite{Ben06}) determined recently abundances for a number of
elements for thin and thick disk stars from the solar neighbourhood. From the
elements considered in that papers the following are relevant for our purposes:
C, O, Na, Mg, Al, Si, Ca, Ti, Cr, Fe, Ni, and Zn.

First we consider from this sample the stars with a [Fe/H] ratio within
$\pm0.05$ of the solar value. There are 6 stars which satisfy this condition
and Table \ref{TabElAbBen05} shows the average abundances $a$ of the above
elements and the average scattering $\sigma_{*}$ of the abundances around the
mean value. For comparison the table also shows the random errors of the
abundance determinations from stellar spectra as given by the authors. These
abundances are surprisingly close to the Solar System abundances, though the
general metallicity is somewhat higher. If the range of metallicities is
increased to $\pm0.1$ of the solar [Fe/H] ratio, the number of stars increases
to 13, but the average values for the abundances are practically unchanged,
i.e., the average abundances given for solar like stars in the table do not
depend substantially on the precise choice of the limit $\Delta\rm[Fe/H]$. Hence
abundances of F~\&~G stars with Solar System metallicity agree rather well with
Solar System abundances as given in Table~\ref{TabAbuSolSy}, that are already
compared in Fig.~\ref{FigPla} with results of our model calculation. Our model
therefore reproduces reasonably well the observed abundances of solar metallicity
stars at the solar cycle.

Second we choose from the sample of Bensby et al. (\cite{Ben05}) and Bensby \&
Feltzing (\cite{Ben06}) the thin disk stars with ages less than 1 Gyrs. Despite
the large uncertainties of such age determinations it seems likely that these
stars belong to the most recently born stars of the sample of thin disk stars.
Their abundances therefore should sample the abundance of the ISM in the solar
vicinity during the last, e.g., 1 \dots\ 2 Gyrs or so. The average abundances
of the elements determined by Bensby et al. (\cite{Ben05}) and Bensby \&
Feltzing (\cite{Ben06}) for these stars are given in Table~\ref{TabElAbBen05}.
Extending the sample to stars with an age less than 2 Gyrs does not result in
significant changes of the average abundances, i.e., the results do not depend
on the precise choice of the age limit. The typical metallicity $Z$ of the
`present' ISM determined in this way (the contribution of N and Ne to $Z$ is
estimated) is slightly higher than the Solar System metallicity, as one may
expect from ongoing element synthesis. Our model results for the present day ISM
abundances are compared in Fig.~\ref{FigAbuFGB} with the observed abundances of
F \& G stars formed within the last Gyr given in Table \ref{TabElAbBen05}. Our
model results for the present ISM are also in good agreement with observations.

\subsubsection{Abundances of B dwarfs}

Stars of early spectral type B have short lifetimes. Therefore they sample
abundances from the present day thin disk. Abundances have been determined in
particular for B stars in stellar clusters and we show in Table
\ref{TabElAbBen05} average abundances taken from the literature for B dwarfs in
stellar cluster with galactocentric distances from a range of $\pm2$ kpc around
the solar cycle. Despite the rather heterogeneous observational material the
scattering of observed abundances around the mean is moderate, i.e., element
abundances in the ring $8\pm2$ kpc around the galactic centre seem to be quite
homogeneous. The average abundances and, thus, the metallicity $Z$, are
typically slightly less than the present-day abundances found from F and G
stars (see Fig.~\ref{FigAbuFGB}), as it is also found by Sofia \& Meyer
(\cite{Sof01}). Figure~\ref{FigAbuFGB} compares our calculated abundances for
the present day ISM with the abundances of B dwarf; the agreement again is
reasonable, but compared to the case of F \& G stars it is worse since abundances
of B dwarfs are lower than of F \& G stars.

\subsubsection{Abundances of H\,II regions}

Abundance determinations from H\,II regions around massive stars give also some
information on the element abundances of the present day interstellar medium,
but only a small number of elements can be analysed. A serious problem is that
the refractory elements are in part or even completely condensed into dust
particles, but the degree of depletion usually cannot be determined with a
sufficient degree of reliability, i.e., the total element abundance remains
uncertain. Therefore we consider here only one case, the Orion nebula, since
the abundances in this nebula should reflect element abundances in the present
day ISM in the solar vicinity. Esteban et al. (\cite{Est04}) determined for the
Orion nebula accurate element abundances for some elements. Table 
\ref{TabElAbBen05} gives their results for some elements which include already
their estimated corrections with respect to dust depletion. These abundances
are almost the same as for young F \& G stars. For the other refractory
elements the degree of depletion is high (Shuping \& Snow \cite{Shu97}) and 
they cannot be used for comparison. Our model is also in accord with these
abundances.

The observational tests show that the rather simple type of model for the
galactic chemical evolution considered here is already suited to describe the
global properties of the disk evolution and in  particular to describe the
evolution of element abundances with sufficient accuracy that it may be taken as
a basis for model calculations of the evolution of the dust content.

\begin{figure}[t]
\centerline{
\resizebox{\hsize}{!}{\includegraphics{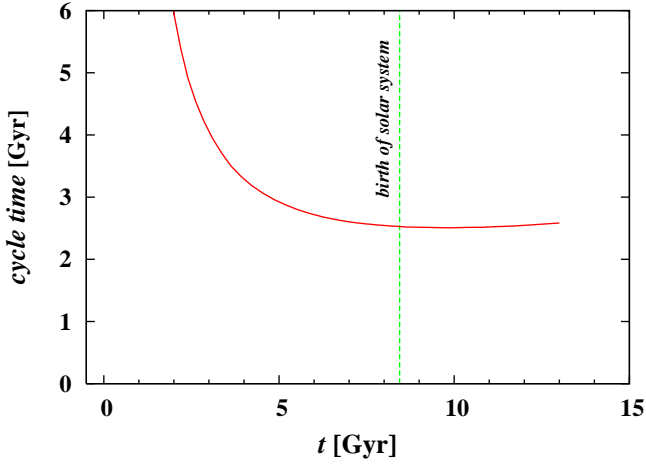}}
}
\caption{Characteristic cycle time of matter between the interstellar medium and
stars at the solar cycle
}
\label{FigCycleTime}
\end{figure}

\subsection{Cycle time for the interstellar matter}
\label{SectCycleT}

In the context of dust evolution an important quantity is the timescale for
conversion of interstellar matter into stars. This is given by
\begin{equation}
\tau_{\mbox{\scriptsize cycle}}={\Sigma_{\rm ISM}\over B}\,.
\end{equation}
This quantity is shown in Fig.~\ref{FigCycleTime}. At the time of formation of
the Solar System the cycle time of matter between the interstellar medium and
stars was about 2.5 Gyrs.


\section{Production rates for stardust}
\label{SectDuEvol}

In the following we study the evolution of the interstellar dust in the Milky
Ways disk within the frame of the simple one-zone approximation for galactic
disk evolution described above. The model follows the principles of the model
of Dwek (\cite{Dwe98}), which was the first one that coupled dust evolution
consistently with a full model for the chemical evolution of the Milky Way. The
present model calculation concentrates on the most abundant dust components
formed in stars that are found also as presolar dust grains in meteorites:
\begin{itemize}

\item silicate dust

\item carbon dust

\item silicon carbide.

\end{itemize}
It also considers
\begin{itemize}
\item solid iron
\end{itemize}
though that is not yet identified as a presolar dust species. For theoretical
reasons it should, however, be an abundant dust species produced in stellar
outflows.

\subsection{Evolution model for the interstellar dust}

In our evolution model we consider a number of different dust species, denoted
by an index $j$. 

First we discriminate between dust coming from different types of parent stars.
Even if the chemical composition of a certain dust species formed in outflows
or ejecta of different stellar types is the same, the individual grains of this
dust species are carriers of the isotopic anomalies corresponding to the
particular nuclear processes operating in their parent stars. If they are
investigated in the laboratory as presolar dust grains, one can, at least in
principle, identify for every grain its formation site. This makes it desirable
to count dust species from different types of stars with the same overall
chemical composition but with different types of isotopic anomalies as different
species~$j$. Presently we choose a not too fine division into stellar types and
consider three different kinds of stellar sources: Supernovae of type II,
supernovae of type Ia, and AGB stars. Supernovae of type II and AGB stars can
form all of the four chemically different types of dust considered in our model,
while supernovae of type Ia probably can form only iron dust (if ever). Hence we
consider in our model nine different kinds of stardust coming from three
different types of parent stars.

\begin{table}

\caption{Dust species considered in the model calculation}
\begin{tabular}{l@{\hspace{1.cm}}c@{\hspace{.5cm}}c@{\hspace{.8cm}}c
@{\hspace{.9cm}}c}
          & silicates & carbon & SiC  & iron \\
\noalign{\medskip}
AGB stars & $\surd$ & $\surd$ & $\surd$ & $\surd$ \\
\noalign{\smallskip}
SN II     &  $\surd$ & $\surd$ & $\surd$ & $\surd$ \\
\noalign{\smallskip}
SN Ia     & --- & --- & --- & $\surd$ \\
\noalign{\smallskip}
ISM       &  $\surd$ &  $\surd$ & --- & $\surd$  
\end{tabular}
\label{TabDuSp}
\end{table}

Second we consider dust formed in the interstellar medium itself. From the
element abundances in the interstellar medium one expects that silicate dust
can be formed. Observations of the interstellar dust indicate, that also
carbon dust can be formed in certain regions of the interstellar medium. It
seems unlikely however, that SiC dust can be formed, since this requires a
carbon rich environment, which is not encountered in interstellar space. Iron
dust may also be formed in the ISM, though this element is probably mainly
consumed by silicate formation. Hence we consider in our model three kinds of
dust formed in the interstellar medium: silicate, iron, and carbon dust.

Totally, our model considers the twelve different kinds of dust from stellar
sources and the interstellar medium given in Table~\ref{TabDuSp}.

We describe the abundance of each dust component~$j$ in the interstellar medium
by its surface mass density $\Sigma_{j,\rm d}$. The evolution of the surface
density is determined by the equation
\begin{equation}
{{\rm d}\,\Sigma_{j,\rm d}\over{\rm d}\,t}=
-{\Sigma_{j,\rm d}\over\Sigma_{\rm ISM}}\,B+
\sum_lR_{j,l,\rm d}-L_{j,\rm d}+G_{j,\rm d}\,.
\label{EqEvolDu}
\end{equation}
The first term on the rhs. describes the loss of dust from the interstellar
medium by star formation. It is assumed that only the dust content of the matter
that is converted into stars is destroyed and that no additional dust is
destroyed during the course of this process. Also no return of freshly
formed dust from protostellar disks by winds or jets is assumed to occur, though
this has been speculated to be possibly important (Tielens \cite{Tie03}). The
second term on the rhs. describes the amount of dust of kind $j$ injected by
stars of type $l$ into the interstellar medium. The third term on the rhs.
describes the losses, the destruction of dust of kind $j$ in the interstellar
medium, mainly by supernova shocks, and the last term describes the gain,
the formation of dust of kind $j$ in the interstellar medium by growth processes
in molecular clouds.

From equations (\ref{EqEvolDu}) one calculates the surface mass density of the
different dust species $j$. Additionally one has the set of equations for the
total surface densities $\Sigma_i$ of each element $i$ in the ISM. The surface
density $\Sigma_{i,\rm g}$ of each element in the gas phase of the ISM then
follows as the difference between its total surface density $\Sigma_i$ and the
sum of the contributions of all dust species containing that element
\begin{equation}
\Sigma_{i,\rm g}=\Sigma_i-\sum_j\nu_{ij}{A_i\over A_{j,\rm d}}\Sigma_{j,\rm d}\,.
\label{GasAbund}
\end{equation}
Here $\nu_{ij}$ is the number of atoms of element $i$ in one formula
unit\footnote{The formula unit is the fictitious molecular group in the solid
corresponding to the chemical formula of the condensed phase} of dust species
$j$, $A_i$ and $A_{j,\rm d}$ are the atomic weights of element $i$ and of one formula
unit of dust species $j$, respectively, and the summation is over all dust
species containing element $i$.

\begin{figure*}

\resizebox{.45\hsize}{!}{\includegraphics{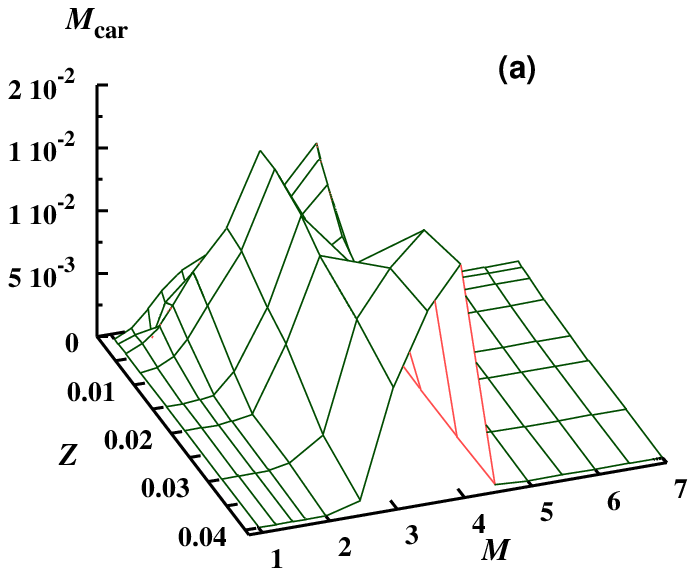}}
\hfill
\resizebox{.45\hsize}{!}{\includegraphics{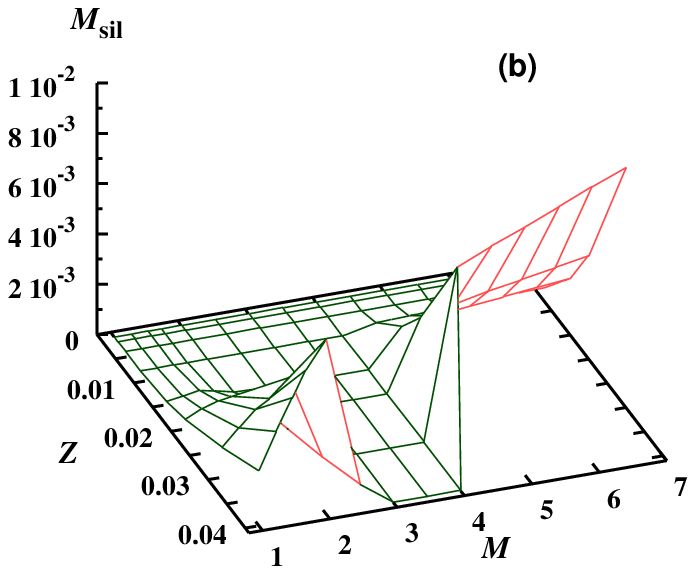}}

\resizebox{.45\hsize}{!}{\includegraphics{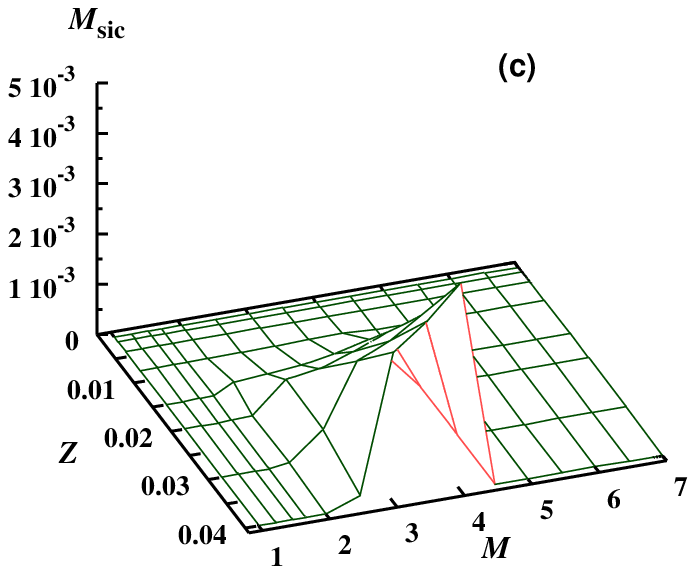}}
\hfill
\resizebox{.45\hsize}{!}{\includegraphics{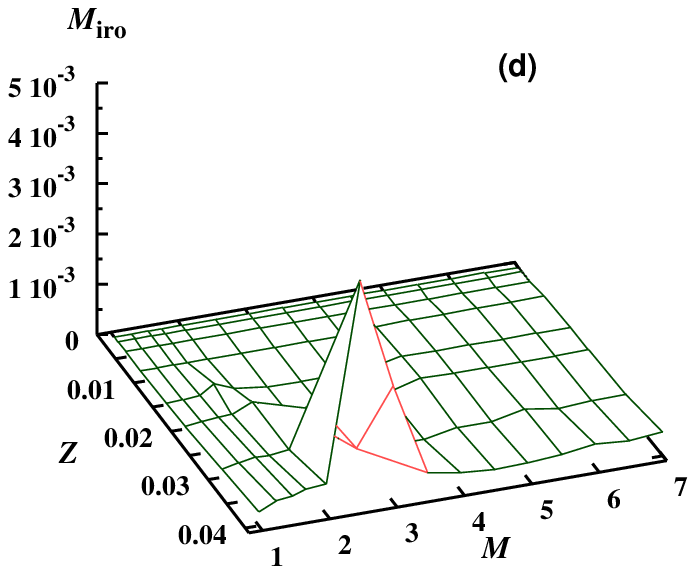}}

\caption{Dependence of the dust masses returned by single
AGB stars for the four main kinds of dust species (silicates, carbon, silicon
carbide, and iron) on metallicity $Z$ and initial stellar mass $M_*$. All
masses are in units of M$_{\sun}$. Data from Ferrarotti \& Gail (\cite{Fer06})
with some additional models}
\label{FigAGBDust}
\end{figure*}

In the following we describe the details of the processes relevant for the
evolution of interstellar dust abundances and how they are implemented in the
model.
 
\subsection{Dust return by AGB stars}
\label{SectAGBDustForm}

The main factories of dust in space are low and intermediate mass stars in the 
AGB stage of their evolution. These are stars with initial masses between about
$0.8\,\rm M_{\sun}$ and about $8\,\rm M_{\sun}$, which end their life as White
Dwarfs. The lower mass limit corresponds to that initial mass, for which the
lifetime of a star corresponds to the age of the Milky Way. The upper mass
limit corresponds to stars which finally explode as supernovae and do not evolve
through an AGB phase\footnote{Presently we neglect the possibility that stars
from the region of initial masses $8<M<12\,\rm M_{\sun}$ may become super-AGB
stars (e.g. Gil-Pons \& Garc{\'{\i}}a-Berro \cite{Gil02})}.

The initial element mixture of all stars is oxygen rich in the sense that the
abundance $\epsilon_{\rm O}$ of oxygen exceeds the abundance $\epsilon_{\rm C}$
of carbon. This doesn't change during their whole evolution up to the
thermally pulsing AGB (TP-AGB), despite some abundance changes during first
and second dredge-up on the Red Giant Branch and the early AGB, respectively.
If the third dredge-up process starts operation, the ashes of He burning are
mixed to the convective envelope of the star after each thermal pulse,
increasing stepwise the carbon abundance of the convective envelope, but only
marginally changes its oxygen abundance. The resulting evolution of the
carbon to oxygen ratio  $\epsilon_{\rm C}/\epsilon_{\rm O}$ on the TP-AGB
depends on the initial mass of the stars:

\smallskip\noindent(1)
Low mass stars with initial masses less than about $M=1.5\,\rm M_{\sun}$ loose
their envelope by a stellar wind before the carbon abundance exceeds the oxygen
abundance. Only for low metallicities their envelope becomes carbon rich prior
to complete envelope loss. For very small initial masses probably no
third dredge-up occurs and the stars never become carbon stars.

\smallskip\noindent(2)
Stars from the range of initial masses between about $M=1.5\,\rm M_{\sun}$ and
about $M=4\,\rm M_{\sun}$ increase after a number of thermal pulses their carbon
abundance over the oxygen abundance and become carbon rich. They further evolve
as carbon stars until their envelope is lost by the stellar wind.

\smallskip\noindent(3)
Intermediate mass stars with initial masses between about $M=4\,\rm M_{\sun}$
and about $M=8\,\rm M_{\sun}$ convert the dredged-up carbon rapidly into 
$^{14}$N via the CN-cycle, since the lower part of their convective envelope 
overlaps with the upper edge of the H burning shell. The oxygen, however, is not
affected by this process, and for this, the carbon abundance in the envelope of
these stars is much less than the oxygen abundance. They do not become carbon
stars until most of their envelope is lost and the convection zone of the
remaining envelope no longer overlaps with the H burning shell. For a short
period the stars then also become carbon stars until finally the last portion of
their envelope is removed by the stellar wind.

\smallskip
Depending on the C/O abundance ratio, the stars produce different dust mixtures
in their outflows. Here we are only interested in the dominating dust species,
which are formed from the most abundant elements and concentrate on the 
following four  types of dust: silicates, carbon, silicon carbide, and iron.
The dust production rate over the whole period of AGB evolution for these dust
species was calculated by Ferrarotti \& Gail (\cite{Fer06}).

The silicate dust is only produced during the oxygen rich phase of the stellar
evolution where the stars spectroscopically appear as M stars. Some minor
fractions are also produced during the S star phase where the C/O abundance
ratio is close to unity. The silicates are a mixture of olivine- and 
pyroxene-type amorphous dust and for part of the stars also up to about
15\% of nearly iron free crystalline forsterite and enstatite is observed to
be formed (cf. the review of Molster \& Waters \cite{Mol03}). The present work
does not discriminate between the two types of amorphous silicate dust since for
silicate dust in the interstellar medium it is presently not possible to
discriminate by observations of the dust absorption spectrum unambiguously
between the two different components (cf. the contradictory results in Chiar \&
Tielens \cite{Chi06} and Min et al. \cite{Min07}). Also crystalline Mg-silicates
are not considered since they are not found in the interstellar medium (Kemper,
Vriend \& Tielens \cite{Kem04}), possibly because they are rapidly amorphized
in the ISM after their ejection by interaction with energetic electrons and ions
(cf. Demyk et al. \cite{Dem04}; J\"ager et al. \cite{Jae03}). 

Carbon and silicon carbide dust are produced by AGB stars during their carbon
rich phase of evolution on the AGB where they spectroscopically appear as C
stars.

Iron dust is included in the model calculation, though it has not yet
unambiguously been identified as major dust component in stellar outflows; only
some hints for its existence have been found up to now (e.g. Kemper et al.
\cite{Kem02}). This is because no readily identifiable spectroscopic features
exist for solid iron. Nevertheless, for reasons of element abundances it should
be an abundantly formed dust species in S stars and C stars and to some extent
also in M stars. 

MgS is also observed to be an abundant dust component in many C stars (cf. 
Molster \& Waters \cite{Mol03}), but is not included in the modelling, since
it is not yet clear by which mechanism it can be formed in a stellar outflow.
This, of cause, presently prevents its modelling.

Figure \ref{FigAGBDust} shows the calculated dust production rates for the four
types of dust considered.  In the model calculation of Ferrarotti \& Gail 
(\cite{Fer06}), olivine- and pyroxene-type dust are treated as separate species
but their production rates are added for the reasons mentioned above.
There is a general tendency for the stars either to be a factory mainly for
silicate dust or to be a factory mainly for carbon dust (cf. Fig. 12 of
Ferrarotti \& Gail \cite{Fer06}), because most of the dust formed over the
total lifetime of a star on the AGB is formed during the very last pulse
cycles on the TP-AGB, where mass-loss rates are highest. If the stars are
carbon stars during this phase they mainly produce carbon dust (and SiC),
otherwise they mainly produce silicate dust.  

The carbon dust production, shown in Fig.~\ref{FigAGBDust}a, is dominated by
stars with initial masses between about 1.5 and $4\,\rm M_{\sun}$ and does not
vary much with initial stellar metallicity, since the carbon required for carbon
dust production is synthesized from He by the star itself. Stars with initial
masses $M>4\,\rm M_{\sun}$ also form some carbon dust, but only small amounts
during their very last stage of evolution when hot bottom burning is no more
active. Stars with initial masses $M<1.5\,\rm M_{\sun}$ do not form much carbon
dust because the total mass returned by them on the AGB is quite small or
because they do not suffer third dredge-up events (for very low initial masses)
or a too small number of them. 

The production of the other dust species by AGB stars strongly depends on their
initial metallicity because the required heavy elements ---with the possible
exception of Mg--- are not fabricated by AGB stars but have to be formed
in many preceding stellar generations until their abundances grew to a level
that dust formation becomes possible.

Figure \ref{FigAGBDust}b shows the silicate dust production by AGB stars.
The silicate production is efficient for stars from essentially that range of
initial masses where they do not become efficient carbon dust producers, i.e.,
the main contribution comes from stars with initial masses $M<1.5\,\rm M_{\sun}$
or $M>4\,\rm M_{\sun}$. But also in the mass range in between, where the stars
are efficient carbon dust factories, they produce some silicate dust before they
become carbon stars. 
The silicate dust production starts to become efficient only at rather high
metallicities because only then sufficient amounts of Si, Mg, and Fe for
silicate formation are available in the stellar outflows.

Figure \ref{FigAGBDust}c shows the production of silicon carbide dust by AGB
stars. This is produced by carbon stars and therefore its production is
limited to the same range of initial masses as for carbon production. The lack
of available Si, however, also prevents the formation of much SiC in low
metallicity stars.

Figure \ref{FigAGBDust}d shows the production of iron dust by AGB stars.
Iron dust formation seems to be efficient in outflows from AGB stars only at
rather high metallicities which are not encountered in the Milky Way at the
solar cycle, but only close to its centre.

The dust-mass injection rate of dust species $j$ into the interstellar
medium is given by
\begin{eqnarray}
R_{j,l,\rm d}(t,r)&=&\int_{M_{\rm l}}^{M_{\rm WD}}dM\,\Phi(M)
{B(t_{\rm b},r)\over M_{\rm av}}\nonumber\\
&&\hspace{1.5cm} M_{j,\rm ret}(M,Z_{\rm ISM}(t_{\rm b},r))\,.
\end{eqnarray}
The index $l$ here refers to AGB stars. The quantity $\Phi$ is the initial mass
function, described in Sect. \ref{SectIMF}, $M_{\rm av}$ is the average mass of
the stars, given by Eq.~(\ref{AverMass}). $B(t,r)$  and $Z_{\rm ISM}(t,r)$ are
the stellar birthrate and the metallicity of the interstellar medium at instant
$t$ and galactocentric radius $r$, respectively. Both quantities are taken from
the model calculation for the evolution of the galactic disk (see 
Figs.~\ref{FigBirthR} and \ref{FigEvolMet}). The instant 
\begin{equation}
t_{\rm b}=t-\tau\left(M,Z_{\rm ISM}(t_{\rm b},r)\right)
\label{EqBirthTime}
\end{equation}
is the time of birth of a star ending its life at instant $t$. The quantity
$\tau(M,Z)$ is the lifetime of a star with initial mass $M$ and metallicity $Z$.
The metallicity of the stars equals the metallicity $Z_{\rm ISM}$ of the
interstellar medium at their birthtime $t_{\rm b}$, that is taken from our
model for the Milky Ways evolution (cf. Fig.~\ref{FigEvolMet}). Note that
$t_{\rm b}$ is given by a non-linear equation which has to be solved numerically
for each $M$ and $t$. Tables for the mass return in the dust species $j$ by AGB
stars of different initial masses and metallicities are given in Ferrarotti \&
Gail (\cite{Fer06}); some additional models have been calculated for the present
work. The integration is performed over the initial masses of the stars from the
lower limit $M_{\rm l}$, here taken to be 1\,M$_{\sun}$, and the upper limit
$M_{\rm WD}$ up to which stars evolve into White Dwarfs, here taken to be
8\,M$_{\sun}$.

\subsection{Dust production by massive stars}

In principle one has three different processes contributing to the dust return
by massive stars that finally explode as SNe:
\begin{enumerate}

\item Dust formed in the massive cool stellar winds of Red Supergiants, i.e.,
massive stars on the Red Giant Branch. This is relevant only for stars from the
range of initial masses $8\la M\la 40\,\rm M_{\sun}$, since only stars from this
mass range enter the Red Giant stage.

\item Dust formed in massive shells of $1 \dots10\,\rm M_{\sun}$ ejected by
repeated giant eruptions during an LBV-phase, like for instance that observed
in $\eta$ Car (cf. Smith \& Owocki \cite{Smi06a}). This is relevant only for
massive stars from the region of initial masses $M\ga 40\,\rm M_{\sun}$ which
evolve through a LBV phase. Most of the mass ejected by these very massive stars
prior to their SN explosion seems to be ejected in a few such events (Smith 
\cite{Smi06b}) which are accompanied by copious dust formation.

\item Dust formed in the ejected matter after the final supernova explosion.

\end{enumerate}
The dust grains formed in outflows from Red Supergiants and giant eruptions of
very massive stars carry isotopic anomalies resulting from hydrogen burning via
the CNO-cycle, while the dust grains formed after a supernova explosion show the
very different isotopic signatures from heavy element synthesis. Both types
would be clearly distinguishable, if investigated as presolar grains in the
laboratory. Both types should be included as separate types of dust in a model
calculation. 

\subsubsection{Dust formed in pre-supernovae}

The dust formed in stellar winds or ejecta prior to the supernova explosion is
later subjected to the shock wave from the SN explosion. This shock wave
destroys the dust in the swept-up material if the expansion velocity exceeds 150
km\,s$^{-1}$ (e.g. Jones et al. \cite{Jon96}). For a simple estimation of the
importance of this process we consider the case that the blast wave expands into
a medium with constant density. At the end of the adiabatic expansion phase the
radius and the velocity of the shocked region are about (e.g. Shull \& Draine
\cite{Shu87})
\begin{eqnarray}
R_{\rm sh}&=&16.2\ E_{51}^{2/7}n_0^{-3/7} \ \rm pc \\
V_{\rm sh}&=&331\ E_{51}^{1/14}n_0^{1/7}\ \rm km\,s^{-1}\,.
\end{eqnarray}
The transition between the Sedov-Taylor expansion and the subsequent pressure
dominated snowplow phase occurs at
\begin{equation}
t_{ST-PDE}=1.91\times10^4\ E_{51}^{3/14}n_0^{-4/7}\ \rm yr\,.
\label{EqSTPDE}
\end{equation}
Here $E_{51}$ is the explosion energy in units of 10$^{51}$ erg and $n_0$ the
density of the ambient medium in units of 1\,cm$^{-3}$. Since in the snowplow
phase the shock velocity drops rapidly, the dust destruction occurs mainly up
to the end of the adiabatic expansion phase given by Eq. (\ref{EqSTPDE}).

{\it Red Supergiants:}
First we consider the case of Red Supergiants and let $E_{51}=n_0=1$. Typical
expansion velocities of stellar winds of supergiants are $v_{\rm exp}=20\ \rm
km\,s^{-1}$. The wind material requires a time of about $R_{\rm sh}/
v_{\rm exp}=790\,000$ yrs to expand to the distance $R_{\rm sh}$. The shock strength then is sufficient to destroy all dust material ejected during a period of
$7.9\times10^5-t_{ST-PDE}\approx 7.7\times10^5$ yrs before the SN explosion.
This is of the order of the evolution time on the Red Giant branch (e.g.
Schaller et al. \cite{Sch92}). The main period of dust formation of such stars,
however, is much shorter. Mass-loss rates of supergiants during the phase
where they are enshrouded by massive dust shells are of the order of $10^{-4}
\dots10^{-3}\,\rm M_{\sun}\,yr^{-1}$ (e.g. van Loon et al. \cite{vLo99}) and
this phase can last at most about $10^5$ yrs otherwise the stellar envelope
over the He core would be lost completely by the stellar wind prior to
explosion, which is not observed for this mass-range. 

Hence, all dust formed by Red Supergiants is expected to be destroyed by the
shock wave of the subsequent supernova explosion. Even if some dust survives in
some cases, Red Supergiants cannot be important sources for interstellar dust.

{\it Luminous Blue Variables:}
The expansion velocity of the matter from giant eruptions is somewhat higher
than for winds of Red Supergiants (cf. Lamers et al. \cite{Lam01}) and may be
as high as 100 km\,s$^{-1}$. Correspondingly, the supernova shock destroys all
the dust that was ejected by a giant eruption if the supernova explosion follows
within about $2\times10^5$ yrs after the end of the LBV phase. The LBV phase,
however, seems to occur during the first transition from the blue to the red in
the Hertzsprung-Russel diagram (Lamers et al. \cite{Lam01}) and is followed by a
WR-phase which lasts about $3\times10^5\dots10^6$ yrs (Meynet \& Maeder
\cite{Mey05}). If the star finally explodes, the velocity of the shock from the
SN explosion is already too slow for destroying the dust at the instant when it
catches up with the ejected LBV shell. Dust formed in giant eruptions could
therefore be an important source for interstellar dust. Unfortunately, however,
there is presently insufficient information on the dust production by these
objects to include them in the model calculation and dust production by LBVs
therefore cannot be considered in our present model calculation.

Clearly, the real situation is more complex since a supernova explodes into the
matter ejected by the stellar winds of the preceding evolutionary stages (cf.
Dwarkadas \cite{Dwa06} for a brief discussion), or into the hot bubbles of
other supernovae, and the dust ejected by one massive star may be subjected to
the SN blast waves of other massive stars from the same stellar cluster. A more
detailed investigation of the whole problem is required to determine the
survival probability of dust formed by a star prior to its SN explosion.

\subsection{Dust return by supernovae}
\label{SectSNDustForm}

Unfortunately it is presently not definitely known which supernovae do form dust
and in which quantities. Undoubtedly there is some dust formed by supernovae
since presolar dust grains are known which bear the signatures of element
synthesis in supernovae. The abundance of X-grains in the population of presolar
SiC grains, however, is small compared to mainstream SiC grains (cf. Hoppe et al.
\cite{Hop00}; Nittler \& Alexander \cite{Nit03}) which are thought to come from
AGB stars. Dust formation by supernovae, therefore, seems to be an inefficient
process. For supernovae of type Ia, observations even seem to indicate that they
do not form dust at all (Borkowski et al. \cite{Bor06}). From the theoretical
side also little is known about dust condensation in SNe; only a few model
calculations for dust condensation in supernova ejecta are available (Kozasa et
al. \cite{Koz89}; Todini \& Ferrara \cite{Tod01}; Nozawa et al. \cite{Noz03};
Schneider et al. \cite{Schn04}) and they are of a very qualitative nature.

Presently there are no reliable models for dust formation in supernovae
available on which one can base a modelling of the contribution of supernovae to
the interstellar and presolar dust population. Therefore we apply in the
present model calculation the same simplified approach as in Dwek (\cite{Dwe98})
to account for the contribution of supernovae to the dust production in the
Milky Way. It is assumed that supernovae of type II produce all types of dust
considered here and supernovae of type Ia produce only iron dust. The dust
return rate is assumed to be given by the total mass return rate of the key
element required to form a particular kind of dust\footnote{
As key element we usually choose that one of all the elements forming the
considered dust species, for which the quantity $\epsilon/i$ takes the smallest
value. Here $\epsilon$ is the abundance of an element by number relative to H,
and $i$ is the number of atoms of the element in the chemical formula of the
compound. The key element determines the maximum amount of dust material that
can be formed for the considered species}
times some efficiency factor $\eta$. This efficiency factor is simply guessed or
is estimated from observational quantities.

We therefore use the following production rates for the dust species
\begin{eqnarray}
R_{{\rm sil},l,\rm d}(t,r)&=&\eta_{\rm sil,SN\,II}\,R_{\rm Si,SN\,II}(t,r)\,
{A_{\rm sil}\over A_{\rm Si}}
\label{SnProdSil}\\
R_{{\rm car},l,\rm d}(t,r)&=&\eta_{\rm car,SN\,II}\,R_{\rm C,SN\,II}(t,r)\,
{A_{\rm car}\over A_{\rm C}}\\
R_{{\rm sic},l,\rm d}(t,r)&=&\eta_{\rm sic,SN\,II}\,R_{\rm Si,SN\,II}(t,r)\,
{A_{\rm sic}\over A_{\rm Si}}\label{DuProSNIISiC}
\\
R_{{\rm iro},l,\rm d}(t,r)&=&\eta_{\rm iro,SN\,II}\,R_{\rm Fe,SN\,II}(t,r)\,
{A_{\rm iro}\over A_{\rm Fe}}\,.
\label{SnProdFe}
\end{eqnarray}
The index $l$ here refers to supernovae of type II. $R_{\rm Si,SN\,II}$ is the
rate of mass return to the interstellar medium of element Si by all supernovae
of type II, defined by
\begin{eqnarray}
R_{\rm Si,SN\,II}(t,r)&=&\int_{M_{\rm WD}}^{M_{\rm u}}dM\,\Phi(M)
{B(t_{\rm b},r)\over M_{\rm av}}\cdot\nonumber\\
&&\hspace{1.5cm} M_{\rm Si,ret}(M,Z_{\rm ISM}(t_{\rm b},r))\,.
\end{eqnarray}
The quantity $M_{\rm Si,ret}(M,Z)$ is the Si mass returned by a supernova with
initial mass $M$ and metallicity $Z$. The integration is from the lower limit
$M_{\rm WD}$ to the upper limit $M_{\rm u}$, here taken to be 40\,M$_{\sun}$
since the tables for mass-return by supernovae of Woosley \& Weaver
(\cite{Woo95}) and Nomoto et al. (\cite{Nom06}) only extend up to this upper
mass.  The mass return rates for the other elements are defined correspondingly.

The quantities $A_{\rm sil}$, $A_{\rm car}$, $A_{\rm sic}$,
$A_{\rm iro}$ are the atomic weights of the dust species and $A_{\rm Si}$,
$A_{\rm car}$, $A_{\rm Fe}$ the atomic weights of the key elements.

\begin{table}

\caption{Characteristic quantities and numerical coefficients used for 
calculation of grain destruction and grain growth (SiC does not form in the ISM)}

\begin{tabular*}{\hsize}{lcccc}
\hline\hline
\noalign{\smallskip}
 & silicates & carbon & iron & SiC \\
\noalign{\smallskip}
\hline
\noalign{\smallskip}
\\
$\tau_{j,\rm SNR}$ (Gyr) & 0.4 & 0.6 & 0.6 & 0.6 \\
$\eta_{\rm SN\,II}$ & 0.00035 & 0.15 & 0.001 & 0.0003 \\
$\eta_{\rm SN\,Ia}$ & 0.0  & 0.0  & 0.005  & 0.0 \\
key element & Si or Mg & C & Fe & Si \\
Atomic weight $A$ & 121.41 & 12.01 & 55.85  &  \\
bulk density $\rho_{\rm c}$ (g\,cm$^{-3}$) & 3.13  & 2.25 & 7.86 & \\
\noalign{\smallskip}
\hline
\end{tabular*}
\label{TabDustPar}
\end{table}

For supernovae of type Ia the dust production rate is
\begin{eqnarray}
R_{{\rm iro},l,\rm d}(t,r)&=&\eta_{\rm iro,SN\,Ia}\,R_{\rm Fe,SN\,Ia}(t,r)\,
{A_{\rm iro}\over A_{\rm Fe}}\,,
\label{DuProSNIaFe}
\end{eqnarray}
where the index $l$ now refers to supernovae of type Ia. The mass return rate
for iron is
\begin{equation}
R_{\rm Fe,SN\,Ia}(t,r)=M_{\rm Fe,ret}R_{\rm SN\,Ia}(t,r)\,,
\end{equation}
where $R_{\rm SN\,Ia}(t,r)$ is the supernova rate of type Ia, which is taken
from the galactic evolution model (see Fig.~\ref{FigBirthR}). $M_{\rm Fe,ret}$
is the iron mass returned by supernovae of type Ia and is taken from the tables
of Iwamoto (\cite{Iwa99}).

The quantities $\eta_{\rm sil,SN\,II}$, ..., $\eta_{\rm iro,SN\,Ia}$ are the
efficiencies for conversion of the key elements of the different dust species
into dust particles and refer to the amount of dust injected into the
interstellar medium in relation to the total mass of the key element returned
to the interstellar medium. The dust first formed in the expanding SN
ejecta is later overrun by the reverse shock and part of it is destroyed again
(Dwek \cite{Dwe05}). Further, part of the surviving dust may be later destroyed
by blast waves from other supernova explosions in the same stellar cluster. The
efficiencies $\eta$ as they are defined here consider all destruction effects
and may therefore be significantly smaller than the efficiency of the initial
dust condensation.

So far only a few attempts have been made to estimate the condensation efficiency
in SNe by analysing spectroscopic data, resulting in very different dust yields,
from only $5\times10^{-4}$ to $4\times10^{-3}$ for type II SN 1987A (Ercolano et
al. \cite{Erc07}) to 0.12 for SN 2003gd (Sugerman et al. \cite{Sug06}). The
values of $\eta$ for different types of SNe are still unknown and have to be
guessed somehow. The numerical values chosen in this paper are much smaller than
the values assumed in Dwek (\cite{Dwe98}) and are given in 
Table~\ref{TabDustPar}. The values for the efficiencies  $\eta_{\rm sil,SN\,II}$,
$\eta_{\rm sic,SN\,II}$, and $\eta_{\rm car,SN\,II}$ of silicate, SiC, and carbon
dust formation in SN~II, respectively, are estimated from the abundances of
presolar silicate grains from supernovae, of X-type SiC grains, and graphite
grains from supernovae. This is discussed in Sect.~\ref{SectEtaSiC}. The
efficiencies $\eta_{\rm iro,SN\,II}$ and $\eta_{\rm iro,SN\,Ia}$ for iron dust
production in SNe of type II and type Ia, respectively, are arbitrarily set 
to a low non-zero value, but they may well be equal to zero.

\subsection{Dust injection rates for the Milky Way model}

\begin{figure}[t]

\resizebox{\hsize}{!}{\includegraphics{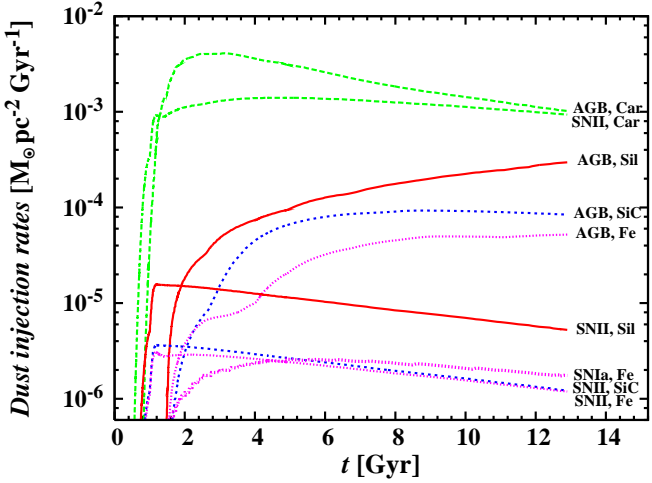}}

\caption{Evolution of the dust injection rates at the solar cycle from different
stellar sources
}
\label{FigDustInjRate}
\end{figure}

Figure \ref{FigDustInjRate} shows the variation with time of the dust injection
rates from stellar sources into the interstellar medium at the solar cycle,
as calculated for our model of the evolution of the Milky Way. The dust 
injection rate in this model is dominated by carbon dust from AGB stars and SNe,
and by silicate dust from AGB stars, except for the very first period before
the first appearance of AGB stars, where dust return from SNe dominates. The SN
injection rates are very uncertain, however, since they depend on the
efficiencies $\eta$ which are only badly known and in this paper are determined
from abundance ratios of presolar dust grains from AGB stars and supernovae.

\section{Dust evolution in the ISM}
\label{SectDestrGro}

\subsection{Dust destruction in the interstellar medium}
\label{SectDustDestruct}

Once condensed in a circumstellar envelope dust grains are driven away from
the star by radiation pressure and are gradually mixed with the ambient ISM. The
analysis of astronomical observations and presolar dust grains from meteorites
indicate a high degree of subsequent dust processing in the ISM. The processes
can be divided in two groups:
\begin{enumerate}
\item the destruction processes (thermal sputtering, evaporation in
high-velocity grain-grain collisions, chemical sputtering), and the growth by
accretion that changes the total dust mass; 
\item
the processes modifying
the grain size distribution or the phase of the grains (shattering in
grain-grain collisions, coagulation, amorphization of crystalline stardust etc.).
\end{enumerate}
Detailed discussions of dust processing in the ISM can be found in Tielens
(\cite{Tie05}), for example. Since in this paper we do not study the grain size
distribution, we are interested in the first group of processes only, i.e., in
the destruction processes and the growth processes in the ISM.

At the stage of grain ejection from the star by radiation pressure, grain
destruction is not important for quiescent mass loss. Typically, drift
velocities for sources with high mass loss are about 1~km~s$^{-1}$ (e.g. Gail \&
Sedlmayr \cite{Gai85}), which is less than the threshold for sputtering of the
dust; grain erosion is therefore unimportant during dust ejection into the ISM
(Woitke et al. \cite{Woi93}, Jones et al. \cite{Jones05}). However, the lower
limit for shattering in grain-grain collisions is about 1-2~km~s$^{-1}$ and
grain fragmentation could be important if there are significant relative
velocities between the grains in stellar envelopes (Jones et al. \cite{Jon96}).
In case of dust condensation in SN explosions freshly formed grains later are
overrun by the reverse shock with velocities of the order of at least $1000\,
{\rm km\ s}^{-1}$. Therefore a significant part of the SN condensates is
destroyed by sputtering in reverse shocks (Dwek \cite{Dwe05}). This is included
in the efficiencies $\eta$ introduced in Sect.~\ref{SectSNDustForm}.

The dominant dust destruction process is dust destruction in the ISM by
sputtering in high-velocity SN shocks ($v > 150\, {\rm km\,s^{-1}}$) (cf. Jones
et al. \cite{Jon96}) that results in injection of atoms into the gas phase by
interaction with impinging energetic ions. This process works almost exclusively
in the warm phase of the interstellar medium, which links the dust destruction
problem in the ISM in principle closely with the multiphase structure of the
ISM. Since we approximate in our present model the ISM by a simple one-phase
model, we describe this process in the approximation which describes the
destruction process in terms of grain lifetimes against destruction by SN
remnants $\tau_{j,\rm SNR}$, i.e., by the average time required to return the
grain mass back to the gas-component of the ISM. The change of surface density
of dust species of kind $j$ per unit time by dust destruction is:
\begin{equation}
L_{j,{\rm d}}=-{\Sigma_{j,{\rm d}}\over\tau_{j,\rm SNR}}\,.
\end{equation}
Since we consider only equations for the total element abundances (gas+dust) in
the ISM and no separate equations for the gas-phase abundance, the mass return
to the gas phase needs no special treatment. This is automatically accounted for
if the gas phase abundances are calculated from Eqs.~(\ref{GasAbund}).

The efficiency of the dust destruction process is determined by the SN II and
SN Ia frequency, the density of the ISM, as well as by the intrinsic properties
of grains such as size, composition, and bond energies. The grain lifetime at
any radius $r$ and instant $t$ can be expressed by the current lifetime at the
solar circle according to Dwek (\cite{Dwe98}) by the approximate scaling law:
\begin{eqnarray}
\tau_{j,\rm SNR}(t,r)&=&\nonumber\\
&&\hskip-.9cm
{\Sigma_{{\rm ISM}}(t,r)\over\Sigma_{{\rm ISM}}(t_{\rm G},
r_{\sun})}\,{R_{\rm SN}(t_{\rm G},r_{\sun})\over R_{\rm SN}(t,r)}\
\tau_{j,\rm SNR}(t_{\rm G},r_{\sun})\,.
\end{eqnarray} 
The total SN rate  $R_{\rm SN}(t,r)$ can be replaced by the stellar birthrate 
$B(t,r)$ with the assumption of spatial and temporal constancy of the IMF. The
dust species considered in our model are listed in Table~\ref{TabDuSp}. For these
dust species we need to know the destruction timescales $\tau_{j,\rm SNR}(t,r)$.

Grain lifetimes $\tau_{j,\rm SNR}(t_{\rm G},r_{\sun})$ have been subject of
many theoretical studies considering thermal sputtering, shattering and models
of SN explosions; the most detailed ones are Jones et al. (\cite{Jon94,Jon96},
see also the references therein). For the current grain lifetimes at the solar
cycle against destruction in all phases of the ISM we take the theoretical
estimates of 0.6 and 0.4 Gyr for carbonaceous and silicate dust, respectively,
from Jones et al. (\cite{Jon96}). Unfortunately they did not present the
corresponding timescales for iron and silicon carbide dust, but their results
for the sputtered dust mass fraction for different shock velocities for iron and
for silicon carbide dust are somewhat higher but similar to carbon dust. We
therefore choose for both the same lifetime of 0.6 Gyr as for carbon dust. The
model results depend only moderately on the value of $\tau_{j,\rm SNR}$ as long
as $\tau_{j,\rm SNR}$ is at least of the order of 0.5 Gyr. This is because for
this and higher values of $\tau_{j,\rm SNR}$ the destruction timescale for dust
becomes roughly comparable with the timescale of conversion of ISM matter into
stars that is of the order of 2.5 Gyrs (cf. Fig.~\ref{FigCycleTime}). In the
opposite case, for timescales $\tau_{j,\rm SNR}$ significantly smaller than
about 0.5 Gyrs, the results depend markedly on the precise value
of $\tau_{j,\rm SNR}$.

The values of $\tau_{j,\rm SNR}(t_{\rm G},r_{\sun})$ used in
the model calculation are also shown in Table~\ref{TabDustPar}.

\subsubsection{Evolution of stardust}

Stardust is only destroyed in the ISM and does not gain mass by accretion of
gas phase material. All material from such grains ejected into the gas phase
rapidly mixes with the existing ISM gas-phase material and the specific isotopic
anomalies carried by the stardust material are lost by mixing together eroded
material of grains from many different kinds of stellar sources. If such
material is later accreted by dust grains in the ISM it shows no isotopic
anomalies. Even if such material grows as mantle material on stardust cores the
differences in isotopic composition between core and mantle survive since dust
grains in the ISM are not expected to ever become hot enough ($>1000$ K) for
sufficient long periods that solid state diffusion smoothes out isotopic
abundance differences between a stardust core and an ISM-grown mantle. Hence any
accreted mantle material can be clearly discriminated (if it could be analyzed
in  the laboratory) from cores originating from stellar sources by showing
isotopic abundances ratios close to Solar System isotopic abundance ratios, even
if the general chemical composition and mineralogical structure of an ISM-grown
mantle material should resemble that of a core with stellar origin. Therefore
we treat the dust species from stellar sources in our model as separate dust
components and omit for these species in Eq.~(\ref{EqEvolDu}) the growth
term; only the destruction and source terms are retained.

\subsection{Dust growth in the interstellar medium}

Dust grains cycle between the cloud and intercloud phase of the ISM on a
timescale of $\simeq3\times10^7$~yr (e.g. Draine \cite{Dra90}; Tielens 
\cite{Tie98}), undergoing destruction in the warm intercloud medium. All
theoretical calculations of grain lifetimes against destruction by SN shocks
agree that they are much shorter than the $\sim$2~Gyr timescale of dust injection
by stars (e.g. Jones et al. \cite{Jon96}; Tielens et al. \cite{Tie05b}). This
requires an efficient mechanism of replenishment of the dust content of the ISM.
Another evidence of dust growth in the ISM is, that gas abundances in the ISM of
major dust forming elements show strong depletion in comparison to solar
abundances (e.g. Savage \& Sembach \cite{Sav96}; Jenkins \cite{Jen05}), which
correlates with the ISM density. The only possible site of grain growth in the
ISM are the dense molecular clouds of the cold phase of the ISM (Draine 
\cite{Dra90}).

It is known that the density of the ISM is not sufficient to allow for formation
of new dust grains, only low temperature accretion of refractory material on
pre-existing stellar grains is possible. The thin mantles accreted in the ISM
are likely more volatile than stellar dust and can be lost more readily during
dust cycling between ISM phases. Besides, in dense molecular clouds the
accretion will be faster, and grains will probably be formed far from
equilibrium, therefore one would expect the grain mantles to be amorphous and
heterogeneous (Jones \cite{Jones05}). Thus, dust accreted in molecular clouds
(the MC-grown dust) has properties different from that of stardust and is
treated in our model as a separate component denoted by an index ISM.

Dust growth in molecular clouds by accretion on existing grains needs to be
considered in our model for silicate and carbon dust, since it is known that
observed depletions of the elements in the ISM are too high as that only
destruction can be active (cf. Jenkins \cite{Jen05} for a recent discussion). The
grains which serve as growth centres for accretion of gas-phase material need not
necessarily be the stardust particles, though these are needed to serve as
initial growth centres for a start-up of the whole process. Also fragments
formed from shattering of MC-grown grains by SN shock waves in the warm
component of the ISM may serve as growth centres for accretion of refractory
elements in the gas phase if mixed into molecular cloud cores.

An unclear case for growth in the ISM is iron dust which might be a component of
the ISM dust mixture if not all Fe is used up by the formation of
magnesium-iron-silicates, in which case it also could grow in the ISM. However,
metallic iron is probably unstable against oxidation in the ISM (Jones 
\cite{Jon90}), while, on the other hand, iron oxides seem not to form a
significant species in the ISM dust mixture (Chiar \& Tielens \cite{Chi06}). We
consider in our model calculation iron as a possible MC-grown dust component
since there are observational indications that not all condensed iron always
resides in silicates (cf. Cartledge et al. \cite{Car06}, their Fig. 10).

How the growth process works in detail is not definitely known. For interstellar
carbon dust it may proceed in the way described in Jenniskens et al.
(\cite{Je93}) as a multistep process, initiated by deposition of ice mantles, and
proceeding via carbonization and polymerization driven by UV irradiation. 
The problem of growth of silicate dust in the ISM has long time remained
unsolved, because the formation of tetrahedral SiO$_4$- structures probably
requires higher temperatures, than the 10-30~K observed in molecular clouds. At
these low temperatures ice mantles are likely formed on the grain surface,
preventing further growth of silicates.
The solution of the problem is possibly provided by intermittent dissipation of
turbulence in molecular clouds (Falgarone et al.\cite{Fal06}). The
large local release of non-thermal energy in the gas by short bursts of turbulent
dissipation has been shown to be able to trigger a specific warm chemistry, which
can be traced by the high abundances of $\rm CH^+$, $\rm H_2O$, and $\rm HCO^+$
observed in diffuse gas. It is shown that signatures of warm chemistry survive in
the gas more then $10^3$yr during chemical and thermal relaxation phase, see
Fig.~10 in Falgarone et al. (\cite{Fal06}). Such local change of the gas
temperature could provide the mechanism for further silicate growth, if the grain
temperature increase is sufficient for ice mantles to evaporate. The latter is
defined by equating energy from collisions with warm gas and the emitted infrared
energy, and thus depends on infrared absorption coefficients of mantle and core
grain material, e.g. it differs noticeably for water and organic ices. 

Our preliminary estimates show that the energy released locally by turbulent
dissipation in molecular clouds is sufficient to evaporate organic ice mantles
from the surface of silicate grains, although detail calculations of temperatures
and residence time in the relaxation phase for grains with different compositions
have to be done to make quantitative estimates. This is a separate problem
important for understanding the physics of dust growth in molecular clouds will
be studied in further papers. 

\subsubsection{Growth of dust grains in molecular clouds}

In calculating the growth rates for the dust species we follow a different
procedure as in Dwek (\cite{Dwe98}). Some modifications are necessary because
(1) we wish to consider specific dust components and not merely the surface
density of dust forming elements residing in some not closer specified dust
components, and (2) since it is assumed that growth of dust is essentially
restricted to molecular clouds (cf. Draine \cite{Dra90}) which relates the dust
growth problem, like the dust destruction problem, closely to the multiphase
structure of the ISM, which has approximately to be taken into account.

It is generally assumed that the growth of dust grains of a specific kind $j$ is
governed by some rate determining reaction step, usually by the addition of that
one of the elements required to form the chemical compound which has the lowest
abundance in the gas phase, and that the rate of addition of all other more
abundant elements adapt to the slowest process. The growth is determined
in this case by some specific key element and a special atomic or molecular
species from the gas phase carrying most or all of this key element, the
\emph{growth species}. The key elements for the condensed phases of interest
are given in Table~\ref{TabDustPar}. The equation for the change of the mass
$m_j$ of a single grain of species $j$ is
\begin{equation}
{{\rm d}\,m_j\over{\rm d}\,t}={\cal A}\,A_jm_{\mbox{\scriptsize\sc amu}}\,{\nu_{j,\rm m}\over
\nu_{j,\rm c}}\,\alpha_j v_{j,\rm th,gr}n_{j,\rm gr}\,.
\end{equation}
Here $n_{j,\rm gr}$ is the particle density of the growth species, $v_{j,\rm th,
gr}$ its thermal velocity, $\alpha_j$ the growth coefficient, $\cal A$ the
surface area of the grain and $A_j$ the atomic weight of one formula unit of the
dust material under consideration. $\nu_{j,\rm m}$ and $\nu_{j,\rm c}$ are the
number of atoms of the key element contained in the growth species and in the
formula unit of the condensed phase, respectively. Evaporation is neglected
since this is not important at the low temperatures in molecular clouds. The
change of the mass density $\rho_j$ of the dust particles in a molecular cloud
is obtained by multiplying the growth equation of single grains by the number
density of grains and the probability distribution of grain radii (assuming
spherical grains) and integrating over all grain radii~$a$. One obtains
\begin{equation}
{{\rm d}\,\rho_j\over{\rm d}\,t}=\alpha_j v_{j,\rm th,gr}n_{j,\rm gr}\,{3V_{1,j}\langle a^2\rangle\over
\langle a^3\rangle}\,{\nu_{j,\rm m}\over\nu_{j,\rm c}}\,\rho_j\,,
\end{equation}
where $V_{1,j}=A_jm_{\mbox{\scriptsize\sc amu}}/\rho_{\rm c}$ is the volume of
one formula unit in the condensed phase, $\rho_{\rm c}$ is the mass density
of the condensed phase, and $\langle\dots\rangle$ denotes averaging with respect
to the size distribution of grains. We define the following average grain radius
\begin{equation}
\langle a\rangle_3^{\phantom{j}}={\langle a^3\rangle\over\langle a^2\rangle}\,.
\end{equation}
For a MRN size distribution (Mathis, Rumpl \& Nordsiek \cite{Mat77}) we have
for instance
\begin{equation}
\langle a\rangle_3^{\phantom{j}}=\sqrt{a_0a_1}\approx 0.035\,\mu\rm m\,,
\label{AvRadMRN}
\end{equation}
where $a_0=0.005\,\mu$m and $a_1=0.25\,\mu$m are the lower and upper limits of
the distribution of grain radii, respectively.

\begin{figure}[t]

\resizebox{\hsize}{!}{\includegraphics{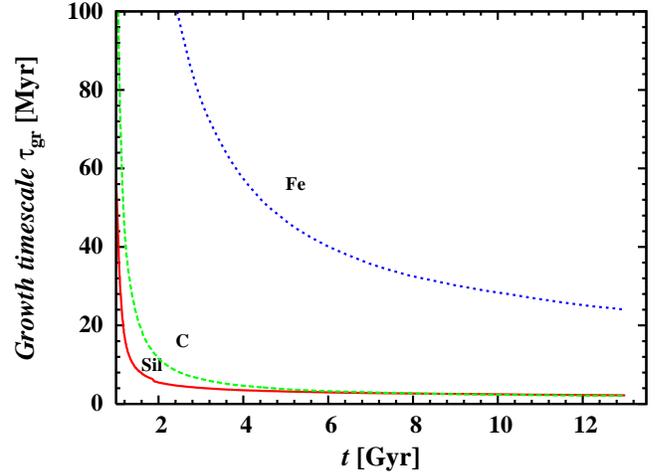}}
\caption{Growth timescale for the dust species growing in molecular clouds
for the Milky Way model at the solar cycle: Silicate dust (full line), carbon
dust (dashed line), iron dust (dotted line)
}
\label{FigGrowtTime}
\end{figure}

The maximum possible particle density of the growth species is
\begin{displaymath}
n_{j,\rm gr,max}={N_{\rm H}\epsilon_j\over\nu_{j,\rm m}}\,,
\end{displaymath}
where $N_{\rm H}$ is the number density of H nuclei in the molecular cloud
(usually equal to $2n_{\rm H_2}$), $\epsilon$ the element abundance of the key
element, possibly lowered by the fraction of this element that is blocked in
some unreactive molecular species. Let $f$ denote the fraction of the key
element already bound in the dust species under consideration, the
\emph{degree of condensation} $f$. The gas-phase density of the growth species
is $\left(1-f\right)n_{j,\rm gr,max}$. Hence we obtain the following equation
for the degree of condensation in the molecular cloud
\begin{equation}
{{\rm d}\,f\over{\rm d}\,t}={1\over\tau_{j,\rm gr}}\,f\left(1-f\right)\,,
\label{EqfCond}
\end{equation}
with
\begin{equation}
{1\over\tau_{j,\rm gr}}=\alpha_j v_{j,\rm th,gr}\,
{3V_{1,j}\over\langle a\rangle_3^{\phantom{j}}}\,{\epsilon\over\nu_{j,\rm c}}
\ N_{\rm H}\,.
\label{TauGrowthDu}
\end{equation}
Numerically we have
\begin{eqnarray}
\tau_{j,\rm gr}&=&46\ {\rm Myr} \times \ \nonumber\\
&&
\quad
\frac {\nu_{j,\rm c}\, {A^{1 \over 2}_{j,\rm m}}} {A_{j,\rm c}}  \left(\rho_{\rm c}\over
3\rm g\,cm^{-3}\right)\left(3.5\,10^{-5}\over\epsilon\right)
\left(103\,{\rm cm}^{-3}\over N_{\rm H}\right)\,,
\end{eqnarray}
where $\tau_{j,\rm gr}$ is evaluated with characteristic values for the physical
variables. The temperature of clouds is assumed to be 10~K, the growth
coefficient $\alpha$ at such low temperatures is assumed to be $\alpha=1$. The
characteristic growth time is generally short compared to the lifetime of
molecular clouds, except at very low metallicity of the ISM. As an example
Fig.~\ref{FigGrowtTime} shows the growth timescale $\tau_{j,\rm gr}$ calculated
from our Milky Way model at the solar cycle for the important dust species.

In principle, the average grain radius $\langle a\rangle_3^{\phantom{j}}$
depends on the degree of condensation $f$ ($a\propto f^{1/3}$ for compact
structures), but we neglect this weak dependence. In this case the equation for
$f$ can immediately be integrated with the result
\begin{equation}
f(t)={f_0{\rm e}^{\,t/\tau_{\rm gr}}\over1-f_0+f_0{\rm e}^{\,t/\tau_{\rm gr}}}
\,.
\label{EvolDegrCond}
\end{equation}
Here $f_0$ is the initial degree of condensation at $t=0$. For $t\gg
\tau_{\rm gr}$ the degree of condensation approaches $f=1$.

\subsubsection{Source term for dust production}

Molecular clouds form in the interstellar medium by instabilities, mainly during
the compression of ISM material in the snowplow phase of SN shocks. They
disappear within a rather short time if active star formation starts and winds
of massive stars and expanding supernova bubbles disperse the clouds. For the
average lifetime of molecular clouds we take an observationally and
theoretically motivated value of $\approx 1\times10^7$ yrs (Leisawitz et al. 
\cite{Lei89}; Williams \& McKee \cite{Wil97}; Matzner \cite{Mat02}; Krumholz 
\& McKee \cite{Krum06}; Blitz et al. \cite{Bli07}). This value for the
lifetime is somewhat shorter than that used in Tielens (\cite{Tie98}) in his
model of dust growth in clouds, but seems to be more appropriate for the most
massive clouds, which contain nearly all of the ISM mass in clouds. The
lifetimes of the clouds equals the characteristic timescale  $\tau_{\rm exch}$
by which matter is exchanged between clouds and the remaining ISM.

At the instant of cloud formation the clouds inherit the dust content of the
interstellar medium outside of clouds. The dust content of the matter outside
of dense clouds is lower than within clouds since dust destruction processes
operate in this material, while in clouds the dust grows by accreting not yet
condensed refractory elements. In fact, except if the metallicity of the ISM is
very low, the growth timescale is much shorter than the lifetime of the cloud
and the condensation of the refractory elements runs into completion before the
cloud disappears.

Let the initial degree of condensation of the key element for some dust species
be $f_0$. If after a period $t$ a cloud is rapidly dispersed, the degree of
condensation in the matter returned to the ISM material outside clouds is equal
to the value given by Eq.~(\ref{EvolDegrCond}). The effective dust mass return
for species $j$ by a molecular cloud is then
\begin{equation}
R_{j,\rm cloud}=\left(f(t)-f_0\right)\,X_{j,max}\ M_{\rm cloud}\,,
\label{ClDustProd}
\end{equation}
where
\begin{equation}
X_{j,\rm max}={A_j\epsilon\over(1+4\epsilon_{\rm He})\nu_{j,\rm c}}
\label{DefCondMax}
\end{equation}
is the maximum possible mass-fraction of the dust species in the material of
the molecular cloud and $M_{\rm cloud}$ is the cloud mass. In principle one has
to observe that some fraction of the cloud mass is converted into stars and not
converted into other phases of the ISM. Since we describe the effect of dust
consumption by star formation within the frame of our approximation by a
separate term in  Eq.~(\ref{EqEvolDu}), this process needs not to be accounted
for in Eq.~(\ref{ClDustProd}).

Equation (\ref{ClDustProd}) has to be multiplied by the probability $P(t)$ that
the cloud is destroyed at some instant within the period between $t$ and 
$t+{\rm d}t$
\begin{equation}
P(t)={1\over\tau_{\rm exch}}{\rm e}^{-t/\tau_{\rm exch}}\,,
\end{equation}
and integrated over $t$. Here it is assumed that the cloud destruction occurs at
random with a mean lifetime $\tau_{\rm exch}$. Finally, averaging with respect
to the mass spectrum of clouds and multiplying with the surface number density
of clouds, one obtains for the mass return rate of MC-grown dust per unit time
and unit area of the galactic disk
\begin{equation}
G_{j,\rm d}={1\over\tau_{\rm exch}}\left(f_{j,\rm ret}-f_{j,0}\right)
X_{j,\rm max}\Sigma_{\rm cloud}\,,
\label{DustRetMC}
\end{equation}
where $\Sigma_{\rm cloud}$ is the surface mass density of clouds and the average
degree of condensation on cloud dispersal is
\begin{equation}
f_{j,\rm ret}={1\over\tau_{j,\rm exch}}\int_0^\infty {\rm d}t\ 
{\rm e}^{-t/\tau_{j,\rm exch}}\,{f_{j,0}{\rm e}^{\,t/\tau{\rm gr}}\over1-f_{j,0}+
f_{j,0}{\rm e}^{\,t/\tau_{j,\rm gr}}}\,.
\label{AvCondDegr}
\end{equation}
The quantity $G_{j,\rm d}$ is the gain term that has to be used in Eq. 
(\ref{EqEvolDu}) for the evolution of the MC-grown dust component $j$.

In principle the evaluation of this term requires to consider a multiphase ISM
where molecular clouds form one of the components. Since we wish to consider the
simpler model of a one-phase ISM we have to cast Eq. (\ref{DustRetMC}) to an
appropriate form for this case. In terms of the mass fraction of clouds in the
ISM $X_{\rm cloud}=\Sigma_{\rm cloud}/\Sigma_{\rm ISM}$ we have
\begin{displaymath}
G_{j,\rm d}={X_{\rm cloud}\over\tau_{\rm exch}}
\left(f_{j,\rm ret}\Sigma_{j,\rm d,max}-\tilde{X}_{j,\rm d}\Sigma_{\rm ISM}
\right)\,,
\end{displaymath}
where
\begin{equation}
\Sigma_{j,\rm d,max}={A_j\over\nu_{i,j} A_i}\Sigma_i
\end{equation}
is the maximum possible surface density of dust of kind $j$ if all material from
the ISM that can be condensed into this dust species is really condensed, and
$\tilde{X}_{j,\rm d}$ is the mass fraction of dust of kind $j$ in that part of
the ISM that is not in clouds. $\Sigma_i$ is the surface mass density of the key
element for species $j$ in the ISM. For the average mass-fraction of dust in the
total ISM we have
\begin{equation}
X_{j,\rm d}=\tilde{X}_{j,\rm d}(1-X_{\rm cloud})+
f_{j,\rm ret}X_{j,\rm d,max}X_{\rm cloud}\,,
\label{AvDustFrac}
\end{equation}
which yields
\begin{displaymath}
\tilde{X}_{j,\rm d}\Sigma_{\rm ISM}={1\over1-X_{\rm cloud}}\Sigma_{j,\rm d}-
{X_{\rm cloud}\over1-X_{\rm cloud}}f_{j,\rm ret}\Sigma_{j,\rm d,max}\,.
\end{displaymath}
It follows
\begin{equation}
G_{j,\rm d}={X_{\rm cloud}\over\tau_{\rm exch}(1-X_{\rm cloud})}\Bigl[\,
f_{j,\rm ret}\Sigma_{j,\rm d,max}-\Sigma_{j,\rm d}\,\Bigr]\,.
\end{equation}
We define the effective exchange time
\begin{equation}
\tau_{\rm exch,eff}=\tau_{\rm exch}{1-X_{\rm cloud}\over X_{\rm cloud}}\,.
\end{equation}
This is much longer than $\tau_{\rm exch}$ since $X_{\rm cloud}\ll1$ and
reflects the fact that it requires many timescales $\tau_{\rm exch}$ to cycle
all ISM material through clouds where it is laden with fresh dust. Our final
result for the dust production term is
\begin{equation}
G_{j,\rm d}={1\over\tau_{\rm exch,eff}}\Bigl[\,f_{j,\rm ret}\Sigma_{j,\rm d,max}-
\Sigma_{j,\rm d}\,\Bigr]\,.
\label{DuProdInCloud}
\end{equation}
This is the appropriate dust production term for MC-grown dust in the approximation
of a one-phase ISM model. In this model the mass-fraction $X_{\rm cloud}$ of
the ISM in clouds is a free parameter which does not follow from the model
calculation but has to be taken from observations. We will use a value of
$X_{\rm cloud}=0.2$ (cf. Tielens \cite{Tie05}), which is appropriate for the
ISM at the solar cycle.

\begin{figure}[t]

\includegraphics[width=\hsize]{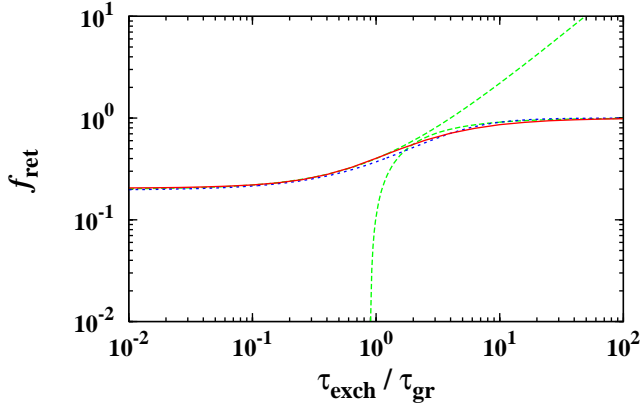}

\caption{Approximation for the variation of the degree of condensation
$f_{\rm ret}$ with $\tau_{\rm gr}/\tau_{\rm exch}$ for $f_0=0.3$. The full line
shows the result of a numerical evaluation of the integral (\ref{AvCondDegr}),
the dashed lines the two limit cases Eqs. (\ref{ApprSlGr}) and (\ref{ApprRapGr}),
and the dotted line the approximation (\ref{ApprAnaGr})
}

\label{FigIntFret}
\end{figure}

\subsubsection{Limit cases}

The degree of condensation $f_{j,\rm ret}$ in the material returned from clouds
at the time of their dispersal depends essentially on the ratio of the
growth timescale $\tau_{\rm gr}$ to the average cloud lifetime $\tau_{\rm exch}$.
If $\tau_{\rm gr}\gg\tau_{\rm exch}$ (slow growth at low metallicities) one
expects that only small amounts of dust are added to the initial dust content,
in the opposite case (rapid growth at normal metallicities) one expects complete
condensation in the returned material. This can be confirmed by calculating the
lowest order terms of a series expansion of the integral in 
Eq. (\ref{AvCondDegr}) for the two limiting cases.

For slow growth ($\tau_{\rm gr}\gg\tau_{\rm exch}$) one introduces $t/
\tau_{\rm exch}$ as integration variable, expands $\exp\,[\,(\tau_{\rm exch}/
\tau_{\rm gr})\,t\,]$ in a series, and integrates term-by-term. The result is
in the linear approximation
\begin{equation}
f_{j,\rm ret}\approx f_0\left(1+{\tau_{\rm exch}\over\tau_{\rm gr}}\right)\,.
\label{ApprSlGr}
\end{equation} 
If this is inserted into Eq.~(\ref{ClDustProd}) one recognizes that 
$\tau_{\rm exch}$ cancels out and, thus, the amount of dust produced during the
residence time of ISM material in the cloud phase is determined by the details
of the growth process.

For rapid growth ($\tau_{\rm exch}\gg\tau_{\rm gr}$) one  introduces $t/
\tau_{\rm gr}$ as integration variable, expands $\exp\,[\,(\tau_{\rm gr}/
\tau_{\rm exch})\,t\,]$ in a series, and integrates term-by-term. The result is
in the linear approximation
\begin{equation}
f_{j,\rm ret}\approx1+{1-f_0\over f_0}\ln(1-f_0)
{\tau_{\rm gr}\over\tau_{\rm exch}}
\label{ApprRapGr}
\end{equation}
(note that $\ln(1-f_0)<0$). If this is inserted into Eq.~(\ref{ClDustProd}) one
finds that the dust production by the clouds is nearly independent of the
details of the growth process within the clouds and is (almost) exclusively
determined by the cycling frequency of ISM material between the clouds and the
other phases of the ISM. The composition of the dust, of cause, is also in this
case determined by the details of growth processes.

The variation of $f_{j,\rm ret}$ with $\tau_{\rm gr}/\tau_{\rm exch}$ in both
limit cases for a value of $f_0=0.3$ are shown in Fig.~\ref{FigIntFret} as
dashed lines together with the result of a numerical evaluation of the integral
(\ref{AvCondDegr}), that is shown as full line. A rather accurate analytic fit
formula for the full range of $\tau_{\rm gr}/\tau_{\rm exch}$ values is
\begin{equation}
f_{j,\rm ret}=\left({1\over
f_{j,0}^2\left(1+(\tau_{\rm exch}/\tau_{\rm gr})\right)^2}
+1\right)^{-{1/2}}\,.
\label{ApprAnaGr}
\end{equation}
This approximation is shown in Fig.~\ref{FigIntFret} as dotted line. The results
for other values of $f_0$ are similar. Only for very small $f_0$ the approximation
becomes somewhat worse in the transition region $\tau_{\rm gr}/\tau_{\rm exch}
\approx1$, but for bigger $f_0$ it becomes even better. For the purpose of
model calculations it suffices to use the approximation~(\ref{ApprAnaGr}).

\subsection{The individual dust species}

Evaluation of the source term Eq.~(\ref{DuProdInCloud}) for dust requires to
calculate the growth time scale $\tau_{j,\rm gr}$, given by 
Eq.~(\ref{TauGrowthDu}), $X_{j,\rm max}$ given by Eq.~(\ref{DefCondMax}), and
the degree of condensation $f_{j,\rm ret}$ in the returned material, which
we calculate from the approximation (\ref{ApprAnaGr}), for all dust species $j$
which are formed by growth in molecular clouds.

The constants required for calculating these quantities are given in
Table~\ref{TabDustPar}. The growth coefficient is assumed to be $\alpha=1$
for all cases since at the low temperatures in dense molecular clouds of about
10 K even the weak attractive van der Waals forces lead to adsorption; the basic
theory for this is discussed, e.g., in Hollenbach \& Salpeter (\cite{Hol70}),
and Watson (\cite{Wat75}). 

For calculating the average $\langle a\rangle_3$ we use in all cases the 
approximation Eq.~(\ref{AvRadMRN}) following from a MRN-size distribution
(Mathis et al. \cite{Mat77}). This is only a crude approximation, but without
attempting to calculate grain size distributions it is hardly possible to fix
this quantity with better accuracy.

The initial value $f_{j,0}$ for calculating $f_{j,ret}$ is given by the degree
of condensation in that part of the ISM matter which is not in clouds, i.e.,
one has
\begin{equation}
f_{j,0}={\tilde X_{j,\rm d}\over X_{j,\rm max}}\,.
\label{DefInValFcond}
\end{equation}
Using this and Eq.~(\ref{ApprAnaGr}) in Eq.~(\ref{AvDustFrac}) yields in
principle a non-linear equation that has to be solved for $f_{j,0}$. For most
purposes it suffices to replace Eq.~(\ref{DefInValFcond}) by the approximation
$f_{j,0}\approx X_{j,\rm d}/X_{j,\rm max}$ since the difference between $\tilde
X_{j,\rm d}$ and $X_{j,\rm d}$ is not very big.

\subsubsection{Silicates}

The silicate dust in the ISM accounts for about one half of the total dust mass
(e.g. Dwek \cite{Dwe05}), but its composition is still a matter of debate.
Studies of silicate composition based on interstellar depletions, modelling of
extinction curve, and in situ measurements of dust in the local ISM give quite
different results, although they all agree on olivine
($\rm[Mg_xFe_{1-x}]_2SiO_4$ with $0<x<1$) and pyroxene
($\rm  Mg_xFe_{1-x}SiO_3$ with $0<x<1$) as major candidates for ISM silicates.
A number of studies of depletions of Mg, Fe, Si and O atoms in the interstellar
gas phase came to the conclusion that observed depletions indicate an
olivine-type stoichiometry of dust in the diffuse ISM (Savage \& Sembach 
\cite{Sav96}, Jones \cite{Jon01}). 
In a recent attempt to fit the silicate features of the interstellar extinction 
curve Min et al. (\cite{Min07}) found that the composition of the ISM silicates
is consistent with a Mg-rich mixture of olivine and pyroxene with a bigger
contribution from pyroxene than from olivine. Fitting of the 9 and 18 $\mu$m
features of the extinction curve shows that, while the 9 $\mu$m feature can be
fitted well by olivine dust, the position and peak strength of 18 $\mu$m feature
is fitted much better with a pyroxene-type stoichiometry (Demyk \cite{Dem99}).

An olivine-pyroxene mixture with a contribution of  more pyroxene than olivine
is therefore chosen for modelling the ISM silicates in the present paper. As a
first approximation we adopt a fixed silicate composition to study silicate dust
production by dust growth in molecular clouds. Modelling of a variable silicate
composition, depending on local growth conditions, is a challenging problem to
be considered in future papers. Let $f_{\rm ol}$ be the (fixed) fraction of the
silicate dust that has olivine stoichiometry; the fraction $1-f_{\rm ol}$ then
has pyroxene stoichiometry.  Assuming the same Mg fraction $x$ for both olivine
and pyroxene in our model, two parameters determine the silicate properties:
$f_{\rm ol}$ and~$x$. 

The total efficiency of dust production by molecular clouds does not show a
significant dependence on the choice of the parameters $f_{\rm ol}$ and $x$.
Variations of the Mg-fraction $x$ change the total dust mass on the level of
10\% at most, but define the silicate-to-iron dust mass ratio. This is due to the
fact, that for the Mg-rich mixtures that are considered here, Mg is the critical
growth species. With decreasing $x$, less Mg is needed for silicate dust growth,
but the total silicate mass increases due to a bigger contribution from the
Fe-bearing component while at the same time less Fe remains for growth of solid
iron. We fix the Mg fraction $x$ to a value of $x=0.8$ by fitting the present-day
silicate-to-carbon dust mass ratio of the model to its observed value of 0.6,
inferred from observations of the infrared emission from the Diffuse Infrared
Background Experiment (Dwek et al. \cite{Dwe97}).

The olivine fraction $f_{\rm ol}$ is chosen such as to reproduce the observed
Mg/Si ratio in dust using the simple relation for a given olivine-pyroxene mixture:
\begin{equation}
f_{\rm ol}={A_{\rm Mg}\over x A_{\rm Si}}-1\,.
\end{equation}
Here $A_{\rm Mg}$ and $A_{\rm Si}$ are observed abundances for the elements Si
and Mg bound in dust (in particles per million hydrogen atoms, ppm). The ratio
${A_{\rm Mg}/A_{\rm Si}}$ equals 1.06 or 1.07 for dust in the diffuse ISM, as
given by Dwek (\cite{Dwe05}) or Whittet (\cite{Whi03}), respectively, which
results in a value of $f_{\rm ol}=0.32$. Although the $A_{\rm Mg}/A_{\rm Si}$
ratio obviously varies in different ISM phases, we use average dust abundances
from the diffuse medium, since this constitutes a significant fraction of the
total ISM mass, and only little is known about the very cold dust in molecular
clouds. Test calculations for different ${A_{\rm Mg}/A_{\rm Si}}$ ratios
available from diffuse ISM studies showed no strong influence on dust masses,
resulting in 4\% change of total dust mass with 10\% decrease of
${A_{\rm Mg}/A_{\rm Si}}$ ratio.

For given silicate composition, the growth species used to calculate the growth
timescale, Eq. (\ref{TauGrowthDu}), is determined by the abundance of the least
abundant species available for dust growth. This is either Si or Mg and we
chose in Eq. (\ref{TauGrowthDu}) 
\begin{eqnarray}
\epsilon &=& 
\cases{\displaystyle
{\Sigma_{\rm Mg}\over24\Sigma_{\rm H}}&  for 
$\displaystyle{\epsilon_{\rm Mg}\over \nu_{{\rm Mg,} c}}<{\epsilon_{\rm Si}\over \nu_{{\rm Si,} c}}$
\cr
\noalign{\bigskip}
\displaystyle{\Sigma_{\rm Si}\over28\Sigma_{\rm H}}& for  
$\displaystyle{\epsilon_{\rm Mg}\over \nu_{{\rm Mg,} c}}<{\epsilon_{\rm Si}\over \nu_{{\rm Si,} c}}$
}\,,
\label{SilGrow}
\end{eqnarray}
where $\nu_{{\rm Si,} c}=1$, $\nu_{{\rm Mg,} c}=1.06$.

\subsubsection{Carbon dust}

The formula unit is the C atom, i.e., one has $\nu_{j,\rm c}=1$. It is assumed
that C is present in the gas phase in molecular clouds predominantly as free
atoms or in a number of molecules bearing one C atom only and that these serve
as growth species. Some fraction $f_{\rm CO}$ of the carbon is blocked in the
CO molecule and is not available for carbon growth. The precise fraction cannot
be fixed without calculating models for the chemistry of the molecular clouds.
Observations indicate a CO abundance in molecular clouds of 20\% \dots\ 40\% of
the C abundance (e.g. Irvine et al. \cite{Irv87}; van Dishoek et al.
\cite{vDi93}; van Dishoek \& Blake \cite{vDi98}). In the calculation we consider
the two cases $\xi_{\rm CO}=0.2$ and $\xi_{\rm CO}=0.4$. The carbon abundance
$\epsilon$ in Eqs.~(\ref{TauGrowthDu}) and (\ref{DefCondMax}) is calculated as
\begin{equation}
\epsilon=(1-\xi_{\rm CO})\,{\Sigma_{\rm C}-\Sigma_{\rm C, sic}\over12\Sigma_{\rm H}}
\,,
\end{equation} 
where $\Sigma_{\rm C, sic}$ is the surface density of C bound in silicon carbide
dust
\begin{equation}
\Sigma_{\rm C, sic}={12\over40}\Sigma_{\rm sic}
\,.
\end{equation}

\begin{figure}[t]

\resizebox{\hsize}{!}{\includegraphics{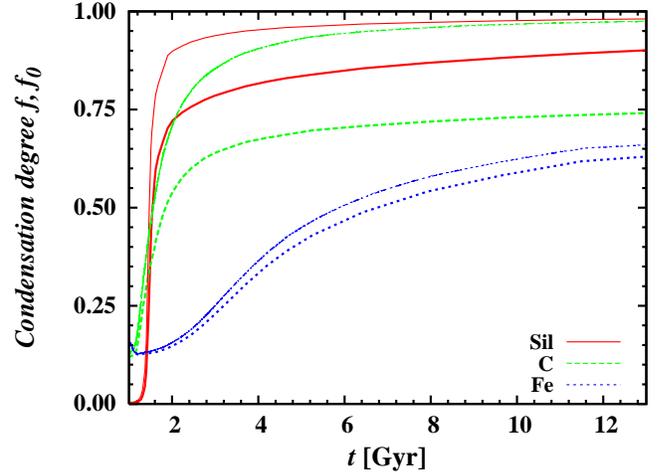}}

\caption{Growth of dust in molecular clouds at the solar cycle. Thick lines show
$f$, the average degree of condensation of the key elements into dust for the
dust species shown at the instant when the molecular clouds are dispersed and
their material is mixed with the other phases of the ISM. Thin lines show $f_0$,
the corresponding degree of condensation at the formation time of clouds. One
always has $f_0<f$ since dust grains grow in molecular clouds and are partially
destroyed again in the ISM outside of clouds until they enter the next cloud.
Growth of iron dust in clouds starts with a significant time delay because of
delayed iron production by SN Ia events. The calculation is for $\xi_{\rm CO}=
0.2$; the result for $\xi_{\rm CO}=0.4$ is not shown because the corresponding
curves are almost the sames.
}
\label{FigDustF0F}
\end{figure}

\subsubsection{Iron dust}

The formula unit is the Fe atom, i.e., one has $\nu_{j,\rm c}=1$. It is assumed
that Fe is present in the gas phase as free atoms, which are the growth species
in this case. The iron abundance $\epsilon$ in Eqs.~(\ref{TauGrowthDu}) and
(\ref{DefCondMax}) is calculated as
\begin{equation}
\epsilon={\Sigma_{\rm Fe}-\Sigma_{\rm Fe, sil}\over56\Sigma_{\rm H}}\,,
\end{equation} 
where $\Sigma_{\rm Fe, sil}$ is the surface density of Fe bound in silicate dust
species.


\begin{figure*}[t]
\sidecaption
\resizebox{11truecm}{!}{\includegraphics{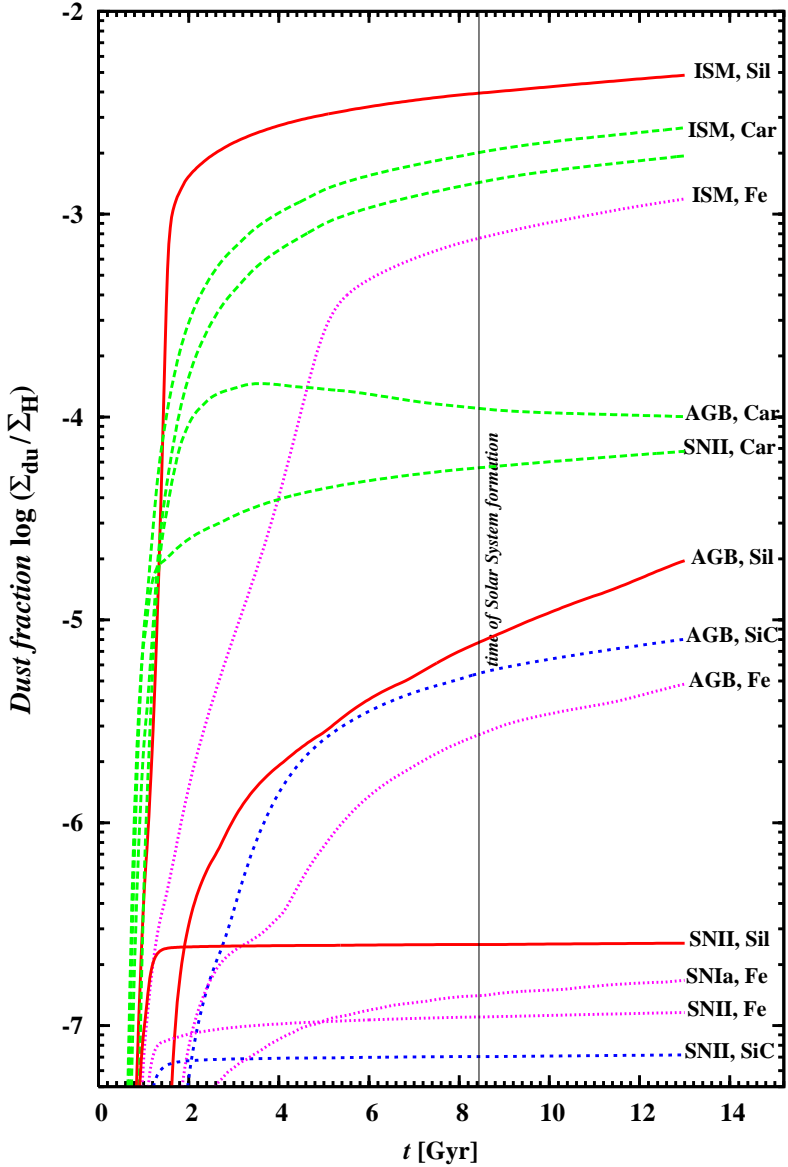}}
\caption{Evolution of the dust mass fraction in the interstellar medium of the
main interstellar dust components and of the stardust species at the solar
cycle. The dust grown in molecular clouds dominates the total dust mass of
the interstellar medium. For carbon dust two results are shown corresponding to
an assumed fraction of 0.2 resp. 0.4 of the carbon in molecular clouds blocked
in the CO molecule. The dust condensed in stellar ejecta (AGB stars, SNe)
has an only small abundance in the ISM. The condensation efficiencies of dust
in supernovae used for the model calculation are given in Tab.~\ref{TabDustPar}.
}
\label{FigEvolDu}
\end{figure*}

\section{Results of the evolution model for the solar cycle}
\label{SectResDuEvol}

\subsection{Evolution of MC-grown dust}

The model for the dust evolution considers silicate dust, carbon dust, and iron
dust as species that grow in dense molecular clouds. The corresponding growth
timescales $\tau_{\rm gr}$ calculated from our Milky Way model at the solar cycle
are shown in Fig.~\ref{FigGrowtTime}.

During the first Gyr of evolution of the galactic disk the metallicity at the
solar cycle is low ([Fe/H]$\la-2$, cf. Fig.~\ref{FigEvolMet}) and the
characteristic growth timescale of dust in clouds exceeds the average lifetime
of dense molecular clouds of about 10 Myr assumed in our model. Only small
amounts of dust are added to the dust content of the interstellar matter during
its cycling through clouds. This can be seen in Fig.~\ref{FigDustF0F} that shows
the evolution of the initial value $f_{j,0}$ of the degree of condensation of the
key elements into dust, defined by Eq.~(\ref{DefInValFcond}), for each of the
dust species $j$, and the average final degrees of condensation $f_j$, calculated
according to Eq.~(\ref{ApprAnaGr}) for the same species, if the clouds are
finally dissolved. Both quantities, $f_{j,0}$ and $f_j$, are calculated during
the course of our model calculation for the evolution of the Milky Way at the
solar cycle. During the first about one Gyr the degree of condensation of
refractory elements into dust increases only marginally by dust growth in
molecular clouds. Therefore, the dust production in the Milky Way is almost
completely determined by dust condensation in the ejecta of stars, and the dust
content of the ISM is determined during this transient phase by dust injection
from stars into the ISM and by dust destruction in the warm phase of the
interstellar medium. Obviously the development would be considerably different
if one has a strong starburst at early times and metallicity becomes already
high before the first AGB stars appear, but this seems not to have happened in
the case of our Milky Way.

Once the metallicity of the ISM has grown to a level of about [Fe/H]$=-2$ some
dust starts to condense during the lifetime of molecular clouds and their dust
content at the instant of their dissolution somewhat exceeds their initial dust
content. From this point on molecular clouds start to contribute to dust
production in the galaxy.

If the metallicity has climbed after more than 2 Gyrs to a level of about
[Fe/H]$=-1$, the degrees of condensation into dust $f_j$ at cloud dispersal
are much higher than the degrees of condensation into dust $f_{j,0}$ at cloud
formation;  in fact, dust growth runs almost into completion during the lifetime
of the clouds. During each cycling step of interstellar matter through clouds
the matter is laden with fresh dust and this dust is mixed into the general ISM
at cloud dispersal. The dust content of the ISM then is essentially determined
by the equilibrium between dust growth in clouds and dust destruction in the
warm phase of the interstellar medium.
 
The degree of condensation $f$ of carbon into carbon dust does not approach
unity (see Fig.~\ref{FigDustF0F}), since it is assumed that 20 to 40\% of the
carbon in molecular clouds forms CO and then is no more available for dust
condensation.

The iron dust abundance evolves somewhat different from that of the silicate and
carbon dust. The main reason is that most of the Fe is produced in SN Ia
explosions and these turn on rather late due to the long lifetime of their low
mass precursor stars. Additionally we assumed in our model that SN Ia explosions
do not start until the metallicity of the precursor stars has rised to
$\rm [Fe/H]\ga-1$ (see Sect.~\ref{SectNuclSynth}). A second reason is that in
our model for dust growth it is assumed that the silicates grown in clouds
contain a certain fraction of iron and the small fraction of iron initially
produced by supernovae is then almost completely consumed by the growth of
silicates with some iron content. This will somewhat change if the iron content
of the silicates is not fixed, as in our present calculation, but will be
determined from growth kinetics.

\subsection{Evolution of dust abundances}

Figure~\ref{FigEvolDu} shows the evolution of the various dust components during
the 13\,Gyrs of evolution of the galactic disk. The dust components with index
`ISM' are the isotopically normal grains grown in the interstellar medium.
Surviving grains from stellar sources are characterized by an index `AGB' or
`SN' if they are from AGB-stars or from supernova ejecta, respectively.

The results depend on the efficiency of dust production by stars, dust
condensation in molecular clouds, and dust destruction rates in the
interstellar medium. The dust production by low and intermediate mass stars on
the AGB is determined from the table of Ferrarotti \& Gail (\cite{Fer06}) and
the dust destruction rate from Jones et al. (\cite{Jon96}). They are probably
not too far from reality. The dust production efficiencies of massive stars are
unknown. One has, however, one piece of information: the abundance ratios of the
presolar dust grains from AGB stars and SNe. We have varied the supernova dust
production efficiencies $\eta$ in Eqs. (\ref{SnProdSil}) \dots\ 
(\ref{SnProdFe}) until the observed abundance ratios for silicate, carbon, and
SiC dust from AGB and SNe sources is reproduced. Details are described in Sect. 
\ref{SectEtaSiC}, the resulting efficiencies are listed in Table 
\ref{TabDustPar}. These efficiencies are very low, probably since they account
also for a number of destruction effects that prevent dust formed in SNe from
escaping into the general ISM.

The dust population of the ISM in this model is dominated by dust grown in
molecular clouds except for the very earliest times, where stardust dominates.
The model shows that presolar dust grains with their isotopic
anomalies revealing the origin of these grains are always a minor component of
the interstellar dust. Most of the dust in the ISM has collected nearly all of
its material from the interstellar gas phase and is isotopically inconspicuous.
If new stars are formed from the ISM containing such a dust mixture, the dust
in their protoplanetary accretion disks contains an only tiny fraction of
presolar dust grains with isotopic anomalies. This fits well with the recent
findings obtained with the nano-SIMS investigations of Interplanetary Dust
Grains (IDPs) by Messenger et al.~(\cite{Mes03}), which show that nearly all of
the silicate grains from cometary nuclei, which should be dominated by
interstellar grains, are isotopically normal\footnote{It is a little bit puzzling
that the STARDUST particles analyzed so far seem to be mainly material from
the Solar System (see Zolensky et al. \cite{Zol06};
McKeegan et al. \cite{McK06})}

The population of stardust grains is dominated by grains from AGB stars because
of the low efficiency of SN dust production. In our model the AGB dust is
dominated by carbon dust; silicate dust and SiC dust is much less abundant. In
meteorites presolar carbon dust in the state of graphite is the least abundant
of these three components (cf. Nguyen et al. \cite{Ngu07}). The discrepancy is
certainly due (i) to the
different survival properties of different kinds of dust material in the Solar
System and the parent bodies of the meteorites, and (ii) the methods of
laboratory investigations applied for different dust grains. This disables
presently a comparison between abundances of different presolar species predicted
by the model and observed in meteorites.

One outstanding feature of the abundance evolution of presolar dust grains is
the rather late appearance of silicate and SiC as compared to carbon grains.
This reflects the fact that AGB stars synthesize the carbon required for soot
formation from He and have not to rely on external sources of heavy elements.
Contrary to this, the Si bearing dust components cannot be formed until
sufficient Si is synthesized in supernova explosions and returned to the ISM
from which subsequent stellar generations inherit the Si required for formation
of Si-bearing species. This needs some time and additionally the precursor stars
of the main sources of Si-bearing dust, the AGB stars, are rather long lived
low-mass stars (cf. Fig.~\ref{FigAGBDust}). Presolar silicate dust grains in the
ISM where a rather new phenomenon at the instant of Solar System formation.

The low abundance of silicate stardust may also provide an explanation for the
lack of crystalline silicate dust in the ISM though a lot of crystalline dust is
injected to the interstellar medium by outflows from AGB stars. Even if there
would be no amorphization processes with energetic electrons and ions
(cf. Demyk et al. \cite{Dem04}; J\"ager et al. \cite{Jae03}), crystalline
silicate dust ($\le 20$\% of the silicate dust injected by AGB stars) would be
too rare compared to amorphous ISM dust to be observable by its absorption
features.

\begin{table}
\caption{SN presolar dust fractions and corresponding derived efficiencies of
dust production.}
\begin{tabular*}{\hsize}{l@{\hspace{1cm}}c@{\hspace{1cm}}c}
	\hline
	\hline
\noalign{\smallskip}
Dust species & Fraction of SN grains & $\eta_{j}$\\
\noalign{\smallskip}
\hline
	\multirow{3}{*}{Silicates}
	& 0.01	& 0.00015  \\
	& 0.03	& 0.00035  \\
	& 0.05 	& 0.0005  \\
\hline
	\multirow{3}{*}{Carbon}
	& 0.1 & 0.04\\
	& 0.3 & 0.15\\
	& 0.5 & 0.20\\
\hline
SiC	& 0.01 & 0.0003\\
\hline
\end{tabular*}
\label{TabRelEtaRat}
\end{table}
\subsection{Efficiency of supernova dust production}
\label{SectEtaSiC}

The dust mass produced in the ejecta of supernovae is not known. Observations
indicate that only small amounts of dust condense and that only part of all
SNe form dust. With the kind of model for dust evolution in the ISM we have
developed, one can try to determine for some dust species an estimate of the
efficiency of dust production by supernovae. This can be done by comparing 
the abundance ratios of supernova dust and AGB dust resulting from the model
calculation with real observed abundance ratios of presolar dust grains with SN
and AGB origin in meteorites. Only silicate, carbon, and SiC dust are presently
suited for this, because only for these dust species the required data for
presolar dust grains are available. Iron dust has not yet been detected as
presolar dust so far and it is unclear whether it really exists.

Such a comparison depends on some assumptions. The first one is that the
production rate of dust by AGB stars is known with significantly better accuracy
than the dust production rate of supernovae. The second basic assumption of a
comparison between these kinds of data is that the fraction of the dust
destroyed between the instant of its incorporation into the just forming Solar
System and the instant of laboratory investigation of presolar dust grains does
not depend on the kind of stellar sources where the dust has formed, but only on
its chemical composition. One has to assume, in other words, that the basic
properties of AGB and SN dust with the same composition, its size spectrum,
its resilience against oxidation and treatment by strong acids, are the same.

\subsubsection{Silicon carbide}

The observed abundance ratio of X-type SiC grains and `mainstream' SiC grains
in the presolar dust population isolated from meteorites is close to 0.01
(Hoppe et al. \cite{Hop00}). Fitting the efficiency $\eta_{\rm sic,SN\,II}$ in 
Eq.~(\ref{DuProSNIISiC}) such that the calculated abundance ratio for SiC from
supernovae of type II and AGB stars agrees with the observed abundance ratio
yields $\eta_{\rm sic,SN\,II}=5\times 10^{-4}$. This ratio seems surprisingly
low, but already the low abundance of X-type SiC grains shows that the
efficiency of SiC dust formation in supernovae is low. The efficiency
$\eta_{\rm sic,SN\,II}$ of SiC dust condensation in supernova determined in this
way is used for the final model calculation and is that given in
Table~\ref{TabDustPar}.

The abundance ratio for the SiC grains refers to grain abundances observed
after isolating the grains from the meteorite matrix by a rather brutal
treatment with oxidizing agencies and by strong acids (cf. Amari et al.  
\cite{Ama94}), but it has been argued by Amari et al. (\cite{Ama94,Ama95a}) that
at most a small fraction of the grain material is lost during this procedure.
On the other hand, the size distribution of SiC grains in Murchison meteorite
found by Daulton et al. (\cite{Dau03}) shows a lack of grains smaller than
0.5 $\mu$m diameter that dominate in circumstellar dust shells (e.g. Jura
\cite{Jur97}), i.e., the small grains are lost already in the ISM or in the
Solar System. If there should be severe systematic differences in the mass
fraction of sub-micron sized grains in the size distributions of SiC grains of
SN and AGB origin, the abundance ratio derived from isolated SiC grains may be
severely misleading, but presently we have no better data.

\subsubsection{Silicate dust}
 
The number of silicate grains from stellar sources detected in meteorites and
interplanetary dust particles is small up to now (Nguyen et al. \cite{Ngu07};
Messenger et al. \cite{Mes05}). Besides the about some 100 silicate grains with
isotopic anomalies attributable to an origin from AGB stars only a single grain
has been detected up to now with isotopic characteristics pointing unambigously
to a SN origin (Messenger et al. \cite{Mes05}). A few more have been detected
that are likely also of SN origin (cf. Vollmer et al. \cite{Vol07}). The small
numbers do not allow to pin down the abundance ratio of silicate dust from the
two possible sources with any reliability, but as a working hypothesis we assume
an abundance ratio of 3\%. Then we can determine an efficiency of silicate dust
formation in supernovae of $\eta_{\rm sil,SN\,II}=3.5\times 10^{-4}$.  This is
the value given in Table~\ref{TabDustPar}. The efficiencies for a somewhat
smaller and bigger abundance ratio are shown for comparison in
Table~\ref{TabRelEtaRat}.

Contrary to the case of SiC the silicate grains are detected by scanning
technics from material that has not been prepared by chemical treatment. There
is therefore no problem to be expected in the sense that part of the grain
population is already destroyed by preparation methods before the particles are
investigated.

\begin{figure}[t]
\resizebox{\hsize}{!}{\includegraphics{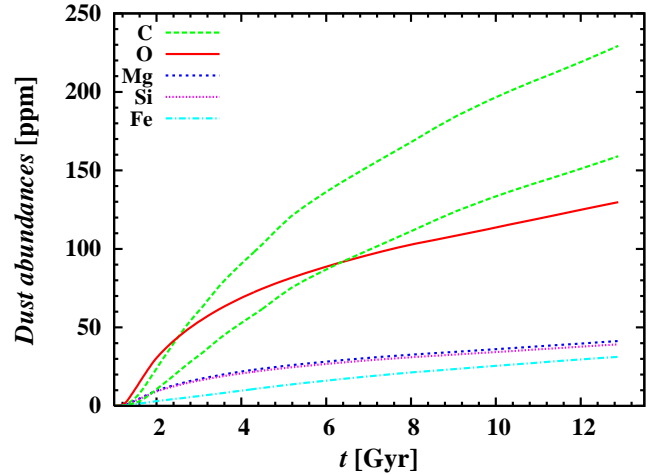}}
\caption{Evolution of abundances in dust per million of hydrogen atoms of the
main dust-forming elements as predicted by the model calculation. For carbon
two lines are shown. The upper one is for the case that a fraction of 
$\xi_{\rm CO}=0.2$ of the carbon is bound in the inreactive molecule CO, the
lower one for the case $\xi_{\rm CO}=0.4$.
}
\label{FigDustPPM}
\end{figure}

\subsubsection{Carbon dust}

The abundance of presolar graphites from supernovae is highly uncertain. 
Chemically separated graphite fractions were further subdivided into low density
separates KE1 and high-density separates KFA1, KFB1 and KFC1 (Amari et al. 
\cite{Ama94}). While many -- though not all -- high density graphites seem to
have an AGB star origin (Croat et al., \cite{Cro05}), low density graphites are
ascribed to supernovae (Hoppe et al. \cite{Hop95}, Amari et al. \cite{Ama95a};
Travaglio et al. \cite{Tra99}), particularly inferred from isotope data of the
low density fraction KE3 (Amari et al. \cite{Ama95b}) which is the coarse grained
($>2\ \mu$m) subgroup making up 70\% of KE1.

If all low density graphites are from supernovae, this would correspond to a 
relative abundance of 67\% (by weight). However, there are significant 
uncertainties which fractions of the various density separates do indeed
correspond to a specific supernova or AGB star signature, so we adopt here an
abundance of $50\pm30$~\% (Hoppe, pers. comm.) and calculate 3 different cases
for 10\%, 30\% and 50\% of all graphites coming from supernovae. For the model
results shown in the figures we assumed a mass-fraction of 30\%.
From this one derives an efficiency of carbon dust formation in supernovae of
$\eta_{\rm sil,SN\,II}=0.15$. This is the value given in Table~\ref{TabDustPar}.
This efficiency is much higher than in the two preceding cases and would mean
that in SN are mainly sources of carbon dust. Efficiencies for a smaller (10\%)
and bigger (50\%) abundance ratio are shown in Table~\ref{TabRelEtaRat} for
comparison.

Presolar graphite grains have mainly size $\ga1\,\mu$m (e.g. Zinner \cite{Zin97})
while for carbon dust grains in circumstellar dust shells around AGB stars one
knows that they have sizes $\la0.1\,\mu$m. Only the small fraction of grains
from a large-size-tail of the size distribution are found in the separates
investigated in the laboratory. If the graphite grains formed in SN ejecta should
have systematically bigger sizes than that formed in AGB-star outflows (there is
however no indication for this), the supernova graphite dust fraction found in
the separates would overestimate the true abundance of graphite grains from
supernovae and our estimated efficiency $\eta_{\rm sil,SN\,II}$ would be too
high.

On the other hand, a much higher condensation efficiency of carbon dust seems
not unlikely since then only the carbon atoms in the carbon layer have to
condense into dust particles and no complicated mixing processes of the 
supernova ejecta are required as for the other dust species.

\begin{figure}[t]

\resizebox{\hsize}{!}{\includegraphics{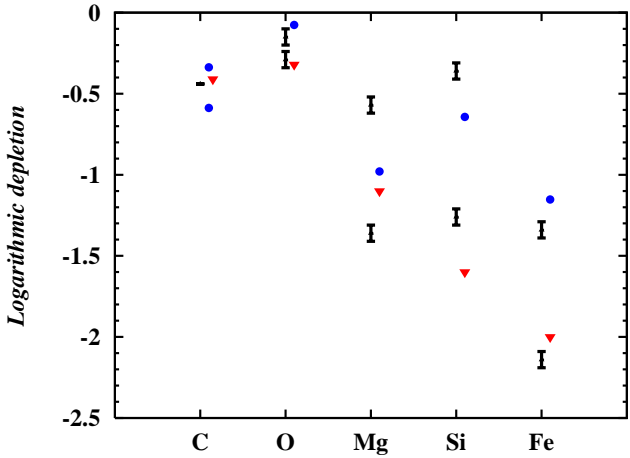}}

\caption{Predicted average depletions of main dust forming elements at present
time are shown with filled circles. Upper and lower points for depletions for 
carbon calculated with CO mass fraction 0.4 and 0.2 correspondingly. Upper and
lower error bars represent observed depletions in  warm and cold diffuse clouds
from Welty et al. (\cite{Wel99}) for C, Si, Fe and Cartledge et al. 
(\cite{Car06}) for O and Mg. Filled triangles marks the average depletions in
diffuse clouds (see Whittet \cite{Whi03} and references therein).}
\label{FigDustDepl}
\end{figure}

\subsubsection{Need for dust accretion in the ISM}

The low efficiency of dust production by supernovae indicated by the rather low
abundance of stardust of SN origin compared to stardust from AGB stars means
that the supernovae cannot contribute substantially to dust in the ISM,
contrary to what is frequently assumed. Therefore it is unavoidable that most of
the dust mass observed to exist in the ISM is formed in the ISM itself and not in
stars.  This has consequences for the dust production in young galaxies with low
metallicity, where only supernovae can be sources for stardust. The high dust
abundances observed in some high-redshift galaxies can, according to our results,
not result from the first generation of SNe but requires already additional
accretion  processes of heavy elements in interstellar clouds. This kind of dust
evolution in young starburst galaxies will be treated in a separate paper.

\subsection{Depletion of elements}

Fig. \ref{FigDustPPM} shows our model results for the abundance evolution of the
main dust forming elements. Since we assumed a fixed composition of ISM
silicates, the ratio between Mg, Si, and O does not change during evolution, but
this is not the case for Fe, which is consumed both by silicate and iron dust
production. In the figure two lines are shown for carbon, corresponding to two
different assumed fractions $\xi_{\rm CO}$ of the carbon in molecular clouds
tied up in the inreactive CO molecule. The upper one corresponds to
$\xi_{\rm CO}=0.2$, the lower one to $\xi_{\rm CO}=0.4$, bracketing typical
observed values (e.g. van Dishoek \& Blake \cite{vDi98}). A higher value of
$\xi_{\rm CO}$ means that less carbon is available for dust formation.

\begin{figure}[t]
\sidecaption
\resizebox{\hsize}{!}{\includegraphics{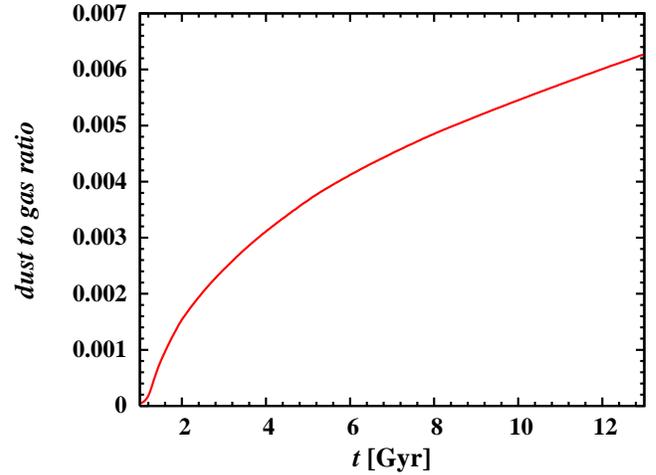}}
\caption{Evolution of the dust-to-gas ratio at the solar cycle as predicted by
the model calculation.
}
\label{FigDustToGas}
\end{figure}

The observed gas phase abundances of elements in diffuse interstellar clouds
indicates various degrees of depletions of many of the dust forming elements
relative to their solar abundances. This is explained as resulting from their
condensation in interstellar dust. The amount of the dust forming elements
locked up in interstellar dust (this is what is shown in Fig.~\ref{FigDustPPM}),
however, cannot reliably be derived from observations. The standard procedure is
to determine instead gas phase abundances of the elements and subtract these
from some kind of `standard' cosmic element abundances in order to determine
how much of each  element is condensed into interstellar dust (cf. Sembach \&
Savage \cite{Sav96} for a review). However, to draw conclusions about the dust
composition from observed depletion patterns one needs to make a decision of
what set of abundances is used as the reference abundances for the elements.
Frequently Solar System abundances, or abundances of nearby F \& G stars or of 
B stars are adopted (cf. Tables \ref{TabAbuSolSy} and \ref{TabElAbBen05}),
resulting in different dust compositions. 

A modelling of the chemical evolution of the Galaxy including dust allows a
study of the evolution of the depletion of the gas abundances by dust
condensation and a comparison of the model with present observed data, since gas
and dust abundances are known from calculations. However, a one phase ISM model
reflects properties of the dust averaged over the different ISM phases,
therefore only a qualitative comparison of depletions is possible. One should
notice that observed depletions are restricted to diffuse clouds, while
molecular clouds are too opaque to be studied in absorption lines.

Our predicted averaged depletions, at present time and at the solar cycle  in
the ISM,  for the 5 main dust forming elements under consideration (C, O, Mg,
Si, and Fe) are shown in  Fig.~\ref{FigDustDepl}. For comparison, observed
depletions in warm and cold diffuse clouds from Welty et al. (\cite{Wel99}) and
Cartledge et al. (\cite{Car06}), and average depletions in diffuse clouds (see 
Whittet \cite{Whi03}) also are shown in the figure. The model calculation
reasonably reproduces the observed values, except for a somewhat low calculated
degree of depletion of Fe.

\begin{figure*}[t]

\hskip 1.2cm
\resizebox{.8\hsize}{!}{\includegraphics{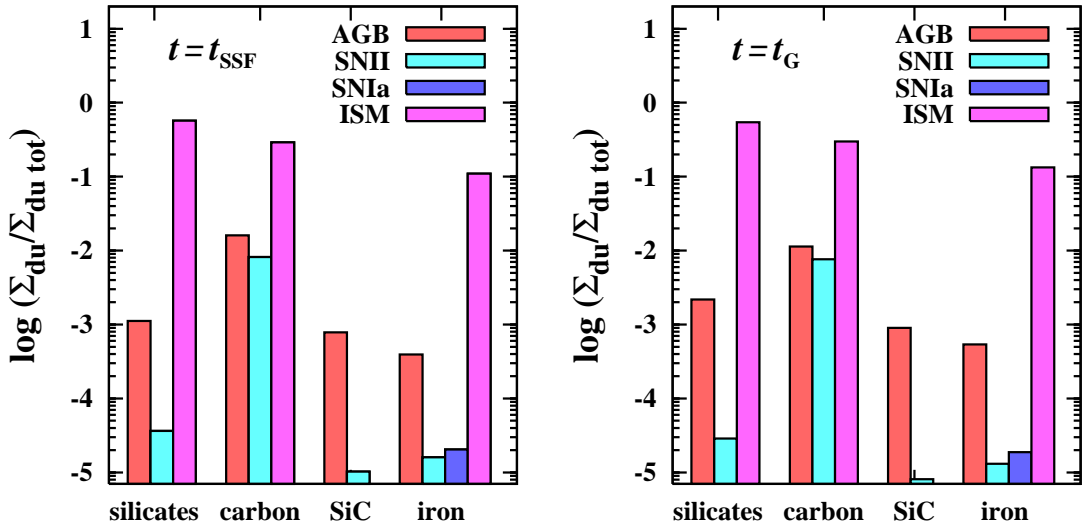}}

\caption{Composition of the interstellar mixture of dust species grown in
molecular clouds, the MC-grown dust, and of the presolar dust species from
AGB stars and supernovae, the stardust, at the solar cycle. {\it Left:} At the
instant of Solar System Formation. {\it Right:} For the present solar
neighbourhood
}
\label{FigCompDu}
\end{figure*}

The degree of depletion of iron cannot be increased by assigning a much longer
destruction timescale $\tau_{j,\rm SNR}$ for iron dust. A model calculation shows
that this does not significantly increase the depletion because the lifetime of
dust grains is in any case limited by the timescale of dust consumption by star
formation, which is about 2.3 Gyr and hence already not really big compared to
the lifetime against destruction by shocks. Also a higher than assumed stability
of Fe-bearing silicate does not help, since then a higher than observed depletion
of Si is to be expected. The main reason for the low degree of depletion in the
model seems to be that the fraction of Fe in the gas phase is not completely
determined by the destruction of Fe-bearing grains by shocks in the warm phase
but to a significant extent also by return of gas-phase Fe by stars, which needs
some time until it is cycled into clouds and depleted from the gas phase by dust
growth processes. In our model the degree of depletion of Fe (and this holds in
principle for all refractory elements) is limited by the rather long time
required for cycling of matter between the ISM matter not in dense clouds and
the matter in dense clouds. The origin of the low degree of depletion of Fe in
the model calculation is presently unclear, but probably bears physical
significance and may indicate that for iron some slow accretion of Fe atoms from
the gas phase into dust is possible also in the warm and/or cold phase of the
ISM.

\subsection{Evolution of the dust to gas ratio}

Figure \ref{FigDustToGas} shows the evolution of the dust-to-gas ratio according
to the model calculation. The hydrogen gas-to-dust mass ratio is approximately
100 for the diffuse ISM averaged over long lines of sight passing through a
number of interstellar clouds (Spitzer \cite{Spi54}). Recent studies of the
hydrogen gas-to-dust ratio in the Local Interstellar Cloud (Kimura et al. 
\cite{Kim04}) also confirms the canonical value from Spitzer (\cite{Spi54}). If
one converts this to a ratio of dust mass to total gas mass, one gets a value of
0.007. The value of the dust-to-gas ratio in our model for the present time ISM
is close to the average value derived from observations. Our model therefore
nicely reproduces the average dust mass fraction of the Milky Way in the solar
vicinity.

\subsection{Dust input into the Solar System}

Figure \ref{FigCompDu} shows the composition of the interstellar dust mixture at
the instant of Solar System formation and the present day composition. 
Numerical values are given in Table~\ref{TabSigDu}. Both mixtures are not
significantly different since the abundances of refractory elements in the ISM
have changed only slightly over the last 4.56 Gyr (cf. Fig.~\ref{FigElRat}).
This dust mixture is clearly dominated by MC-grown dust and contains an only
small fraction of stardust. The stardust is dominated by dust grains from AGB
stars; dust grains with SN origin form only a minor component. The composition
seems to be consistent with models for the composition of the local interstellar
dust (Kimura et al. \cite{Kim03b}; Frisch \cite{Fri06})

\begin{table}
\caption{Surface densities of different dust species at instant of Solar System formation and at present time as predicted by the model calculation.}
\begin{tabular*}{\hsize}{llcc}
	\hline
	\hline
\noalign{\smallskip}
Dust species & Source & $\Sigma_{j{\rm ,d}}(t_{{\rm SSF}}) {\rm [M_{\sun} pc^{-2}]}$ & $\Sigma_{j{\rm,d}}(t_{\rm G}) {\rm [M_{\sun} pc^{-2}]}$\\
\noalign{\smallskip}
\hline
	\multirow{3}{*}{Silicates}
&AGB 	& 5.39$ \cdot 10^{-5}$ & 1.12$ \cdot 10^{-4}$ \\
&SN II 	& 2.51$ \cdot 10^{-6}$ & 2.08$ \cdot 10^{-6}$ \\
&ISM 	& 2.76$ \cdot 10^{-2}$ & 2.76$ \cdot 10^{-2}$ \\
\hline
	\multirow{3}{*}{Carbon}
&AGB	& 7.73$ \cdot 10^{-4}$ & 5.73$ \cdot 10^{-4}$ \\
&SN II	& 7.86$ \cdot 10^{-5}$ & 7.73$ \cdot 10^{-5}$ \\
&ISM 	& 1.43$ \cdot 10^{-2}$ & 1.54$ \cdot 10^{-2}$ \\
\hline
	\multirow{4}{*}{Iron}
&AGB	& 1.89$ \cdot 10^{-5}$ & 2.75$ \cdot 10^{-5}$ \\
&SN II	& 7.72$ \cdot 10^{-7}$ & 6.63$ \cdot 10^{-7}$ \\
&SN Ia	& 9.83$ \cdot 10^{-7}$ & 9.55$ \cdot 10^{-7}$ \\
&ISM 	& 5.31$ \cdot 10^{-3}$ & 6.78$ \cdot 10^{-3}$ \\
\hline
	\multirow{2}{*}{SiC}
&AGB	& 3.80$ \cdot 10^{-5}$ & 4.58$ \cdot 10^{-5}$ \\
&SN II	& 8.21$ \cdot 10^{-7}$ & 6.82$ \cdot 10^{-7}$ \\
\hline
\end{tabular*}
\label{TabSigDu}
\end{table}

The dust mixture at time of Solar System formation is that from which the solid
bodies in our planetary system formed. Relics of this dust mixture can be found
in the Solar System in two types of objects: matrix material of primitive
meteorites and in comets. However, until Solar System bodies form from the dust
component of the matter collapsed from some part of the parent molecular cloud
into the protoplanetary accretion disk, the material underwent a number of
alteration processes. Even the most primitive material in Solar System bodies is
not just unmodified ISM matter. For this reason meteoritic matrix material is
presently not suited for a comparison with the model results, since the
alteration processes on the parent bodies are presently not completely
understood (cf. McSween et al. \cite{McS02}). Material from comets is probably
more suited, and once more detailed results from the STARDUST mission are
available, it may be possible to compare the model predictions for the ISM dust
composition entering the protoplanetary accretion disk of the Solar System with
observations. Presently most of the analyzed particles returned by the STARDUST
mission seem to be material from the solar system (Zolensky et al. \cite{Zol06};
McKeegan et al. \cite{McK06}).

Presently one can only state that the dust mixture inherited by the Solar System 
from its parent molecular cloud and the present day ISM dust mixture in the solar
neighbourhood predicted by our model calculation are roughly in accord with the
dust composition estimated by Pollack et al. (\cite{Pol94}) from observations of
extinction properties of the dust material and considerations on element
abundances, which is presently held for the best estimate of the composition of
the dust material from which the Solar System formed.


\section{Concluding remarks}
\label{SectCR}

In the present paper we have developed a model for calculating the chemical
evolution of of the Milky Way disk and the dust content of the interstellar
medium in a consistent fashion, following partially the methods proposed by 
Tielens (\cite{Tie98}) and in particular by Dwek (\cite{Dwe98}). 

The chemical evolution part of the model for the Milky Ways disk follows mostly
well-established methods. The model is checked with the standard tests applied
to such type of models and reasonably reproduces the observational constraints.
A new aspect is the use of the new tables of Nomoto et al. (\cite{Nom06})
for the heavy element production by massive stars. This gives better agreement
between the calculated evolution of the element abundances of the main dust
forming elements (O, Mg, Si, Fe) and the observed evolution of abundances in the
ISM as witnessed by the variation of atmospheric element abundances of main
sequence G stars with metallicity. In particular the problem with the low Mg
abundances disappears with the new results of Nomoto et al. (\cite{Nom06}). A
good reproduction of the abundance variations of the dust forming elements is
important if one tries to model the interstellar dust mixture.

The model for the evolution of the interstellar dust is based on three essential
elements:

First, the model uses for the dust input to the ISM by low and intermediate mass
stars the results of the model calculations of Ferrarotti \& Gail (\cite{Fer06}),
which combine synthetic AGB evolution models with models for circumstellar dust
shells including dust formation in the stellar wind. These models describe for
the first time consistently the dependence of dust production by AGB stars on
stellar initial mass and metallicity.

Second, the dust production by supernovae is described by a simple
parametrization, that was already applied by Dwek (\cite{Dwe98}). A lot of
information has accumulated over the years  by the efforts of the meteoritic
science community on studying nucleosynthetic processes in stars by analyzing
isotopic abundances in presolar dust grains. As a by-product this provides us
with abundance ratios of presolar dust grains from supernovae and from AGB stars.
This allows for the first time to estimate the efficiency of dust production
by supernovae by fitting calculated abundance ratios of stardust in the Solar
System to observed abundance ratios of presolar dust grains from AGB stars and
SNe. Applying these gauged efficiencies yields the unexpected result, that dust
production by massive stars is not important for the evolution of the ISM dust
component, at least not for Pop II and Pop I metallicities.

Third, a simple approach is developed to include dust growth in molecular clouds
into a model for the evolution of the interstellar dust. This approach is in a
sense incomplete, since the growth of dust (as its destruction) depend heavily
on the phase structure of the interstellar gas (hot, warm, cold), which cannot
be treated adequately within the approach generally used in chemical evolution
models. So some quantities in the dust growth model like the mass fraction of
ISM matter in clouds have to be taken from observations, but we have data for
these quantities only for the present day Milky Way. Since the distribution of
the ISM over the phases is neither spatially nor temporal constant, in a
realistic modelling the phase structure should  be part of the model calculation.
Nevertheless the present approach allows for the first time to study the
evolution of the interstellar dust population, including several stardust
species.

The results obtained with this model are in reasonable accord with, for instance,
the observations of interstellar depletions of refractory elements. Again, one
has the problem that the degree of depletion depends on the phases of the ISM,
which are not adequately considered in the present model. Presently we have
started work to implement our dust model in a chemodynamical evolution code 
(Berczik et al. \cite{Ber03}) which removes the shortcomings of the present
model.


\begin{acknowledgements}
We acknowledge P. Hoppe for some very valuable discussions and comments on
an earlier version of this paper.
This work was supported by the Deutsche Forschungs\-gemeinschaft (DFG),
Sonderforschungsbereich 439 ``Galaxies in the Young Universe''. S. Z.
is supported in part also by the International Max-Planck Research School
(IMPRS) Heidelberg
\end{acknowledgements}


\end{document}